\documentclass[notitlepage,aps,prd,showpacs,superscriptaddress,twocolumn]{revtex4-1}
\usepackage{graphicx}
\usepackage{dcolumn}

\usepackage{amsmath}
\usepackage{amssymb}
\usepackage{bm}
\usepackage{color}
\usepackage[normalem]{ulem}
\usepackage[dvipsnames]{xcolor}
\usepackage{hyperref}
\hypersetup{
colorlinks=true, 
linkcolor=blue,  
citecolor=cyan,  
}
\begin{document}
\title{
Gravitational lensing around Kehagias-Sfetsos compact objects surrounded by plasma}

\author{Sudipta Hensh}
\email{f170656@fpf.slu.cz, sudiptahensh2009@gmail.com}
\affiliation{Institute of Physics and Research Centre of Theoretical Physics \& Astrophysics, Faculty of Philosophy \& Science, Silesian University in Opava,
Bezru\v{c}ovo n\'{a}m\v{e}st\'{i} 13, CZ-74601 Opava, Czech Republic}

\author{Ahmadjon~Abdujabbarov}
\email{ahmadjon@astrin.uz}
\affiliation{Center for Field Theory and Particle Physics and Department of Physics, Fudan University, 200438 Shanghai, China }
\affiliation{Ulugh Beg Astronomical Institute, Astronomicheskaya 33,
	Tashkent 100052, Uzbekistan }
\author{Jan Schee}
\email{jan.schee@fpf.slu.cz}
\affiliation{Institute of Physics and Research Centre of Theoretical Physics \& Astrophysics, Faculty of Philosophy \& Science, Silesian University in Opava,
Bezru\v{c}ovo n\'{a}m\v{e}st\'{i} 13, CZ-74601 Opava, Czech Republic}

\author{Zden\v{e}k Stuchl\'{i}k}
\email{zdenek.stuchlik@fpf.slu.cz}
\affiliation{Institute of Physics and Research Centre of Theoretical Physics \& Astrophysics, Faculty of Philosophy \& Science, Silesian University in Opava,
Bezru\v{c}ovo n\'{a}m\v{e}st\'{i} 13, CZ-74601 Opava, Czech Republic}
\date{\today}
\begin{abstract}

We study the optical properties of the Kehagias-Sfetsos (KS) compact objects, characterized by the ``Ho\v rava'' parameter $\omega_{_{KS}}$, in the presence of plasma, considering its homogeneous or power-law density distribution. The strong effects of both ``Ho\v rava'' parameter $\omega_{_{KS}}$ and plasma on the shadow cast by the KS compact objects are demonstrated. Using the weak field approximation, we investigate the gravitational lensing effect. Strong dependence of the deflection angle of the light on both the ``Hora\v va'' and plasma parameter is explicitly shown. The magnification of image source due to the weak gravitational lensing is given for both the homogeneous and inhomogeneous plasma.

\end{abstract}


\maketitle

\section{Introduction}

Ho\v rava proposed a field-theory approach to quantum gravity inspired by Lifshitz's ideas of solid state physics, based on an anisotropic scaling of space and time~\cite{Horava09a,Horava10}. The Lagrangian of the Ho\v rava theory is Lorentz invariant at low energies, but the invariance is violated at high energies.
Later, within a slightly modified theory, the spherically symmetric, asymptotically flat solution has been found by Kehagias and Sfetsos~\cite{Kehagias09}. The Kehagias-Sfetsos (KS) solution  is compatible with the Minkowski vacuum and includes an extra new parameter $\omega_{_{KS}}$ reflecting the quantum effects. It coincides with the Schwarzschild solution in the limit of large values of $\omega_{_{KS}} M^2$, for the source with total gravitational mass $M$.
Considering $\omega_{_{KS}}$ as an universal constant, one may consider the spacetime to be regulated only by the mass of the object. For the case $\omega_{_{KS}} M^2\geq 1/2$, the solution describes a black hole with  an event horizon, while for $\omega_{_{KS}} M^2<1/2$ it describes a naked singularity. The limits/constraints on $\omega_{_{KS}}$  obtained by using the observational tests do not exclude the existence of the compact objects described by KS solution~\cite{Iorio10,Liu11a,Iorio11}.
For example, the Solar system test gives the limit of $\omega_{_{KS}} > 3.2\times10^{−20}\ {\rm cm}^{−2}$ and implies that the total mass of the object cannot exceed $2.6\times10^4 M_\odot$~\cite{Iorio11}. In the present paper, we are motivated to consider possibility of testing the KS solution using optical properties of the spacetime. The properties of the KS spacetime have been studied by various authors, see, e.g.~\cite{Stuchlik15a, Stuchlik14, Vieira14, Goluchova15}. The particle motion around KS spacetime have been studied in~\cite{Abdujabbarov11a,Enolskii11,Stuchlik14a}.   


One of the basic features of the metric theories of gravity is the gravitational lensing or light deflection effect due to gravitational interaction. It was first discovered by Einstein within the General Relativity and now it is considered as an useful tool to study either source or lens system.
The effect of gravitational lensing is reviewed in~\cite{Synge60,Schneider92,Perlick00,Perlick04}. 
Beside the gravitational force, the plasma surrounding the compact object may also significantly affect the photon motion. The effect of plasma on photon motion in various spacetimes and plasma configurations has been studied by number of authors~\cite{Rogers15,Rogers17,Er18,Rogers17a,Broderick03,Bicak75, Kichenassamy85,Perlick17,Perlick15,Abdujabbarov17,Eiroa12, Kogan10,Tsupko10,Tsupko12,Morozova13,Tsupko14,Kogan17, Hakimov16,Turimov18,Benavides16,Kraniotis14}. 

Recently, image of supermassive black hole in the center of galaxy M87 has been disclosed~\cite{2019ApJ...875L...1E,2019ApJ...875L...2E,2019ApJ...875L...3E,2019ApJ...875L...4E,2019ApJ...875L...5E,2019ApJ...875L...6E}. This observation is due to the Event Horizon Telescope (EHT) based on the very large interferometry (VLBI) technique promise to get deep understanding of the strong gravitational field regime around supermassive black hole (SMBH) and test the theories of gravity. The image of the SMBH or so-called shadow of the black hole has been theoretically studied by many authors \cite{Takahashi05, Hioki09, Bambi09, Bambi10, Bambi12, Amarilla10, Amarilla12, Amarilla13, Abdujabbarov13c, Atamurotov13, Wei13, Atamurotov13b, Bambi15, Ghasemi-Nodehi15, Cunha15, Abdujabbarov15, Atamurotov15a, Ohgami15, Grenzebach15, Mureika17, Abdujabbarov17b, Abdujabbarov16a, Abdujabbarov16b, Mizuno18, Shaikh18b, Kogan17, Perlick17,Schee15, Schee09a,Schee13JCAP, Stuchlik14, Schee09, Stuchlik10, Abdikamalov19, Gott19}. Here we study the effect of the ``Ho\v{r}ava'' parameter on the image of the shadow of the KS compact objects and in the weak-field limit its influence on the gravitational lensing, both in the presence of plasma. 


The paper is organized as follows.
In Sect.~\ref{shadowsec} we introduce the notion of the shadow of the KS compact object in vacuum. In Sect.~~\ref{shadowplas} we investigate the influence of plasma on the shadow of the KS compact object.
Then in Sect.~\ref{photmotion} we review the photon motion around compact object in the presence of plasma. We apply the general formalism to KS spacetime and study the gravitational lensing effect around KS compact object in the presence of plasma. In the next Sect.~\ref{magn} we consider the magnification of image source due to lensing in the presence of plasma. Finally, in Sect.~\ref{Summary} we summarize our results. Throughout the paper we use space-like signature $(-,+,+,+)$, the geometric system of units in which $G = 1= c$ and we restore them when we need to compare our results with observational data. Greek indices run from $0$ to $3$, Latin indices from $1$ to $3$. 
%
\section{Shadow of black hole in vacuum \label{shadowsec}}
\subsection{Kehagias-Sfetsos spacetime}

The metric of the Kehagias-Sfetsos~(KS) spacetime, expressed in the standard Boyer-Lindquist coordinates and geometric units can be written as~\cite{Kehagias09} 
\begin{equation}\label{eq:1}
ds^2 = -f(r)dt^2 + f^{-1}(r)dr^2 + r^2(d\theta^2 + \sin^2\theta d\varphi^2) \ ,
\end{equation}
where the lapse function reads 
\begin{equation}\label{eq:2}
f(r) = 1 + r^2\omega_{_{KS}} \left[1-\left(1+\frac{4M}{\omega_{_{KS}}r^3}\right)^{1/2}\right] \ ,
\end{equation}
and $\omega_{_{KS}}$ is the ``Ho\v{r}ava'' parameter.
\subsection{Equations of geodesic motion}
We treat the equations of motion by following the Hamilton-Jacobi formalism. The Hamilton-Jacobi equation reads 
\begin{equation}\label{eq:3}
\frac{\partial S}{\partial \lambda}=\frac{1}{2} g^{\mu \nu} \frac{\partial S}{\partial x^\mu} \frac{\partial S}{\partial x^\nu} \ ,
\end{equation}
where $S$ is the Hamilton-Jacobi action, and $\lambda$ is the affine~parameter that changes along the geodesic. The four momentum of a test particle is related with the action as $p_\alpha=\partial S/\partial x^\alpha$ \ . Because of the symmetries of the KS spacetime, we can apply separation of the variables and the action can be written as 
\begin{equation}\label{eq:4}
S=-\frac{1}{2} m^2 \lambda-Et+L\phi+S_r(r)+S_\theta(\theta) \ ,
\end{equation}
where $m$ is the mass of the test particle~($m=0$ in the case we are dealing with, i.e. for photons), $E$ is the energy of the particle (photon), $L$ is the axial angular momentum of the particle (photon), $S_r(r)$ is a function of $r$ and $S_\theta(\theta)$ is a function of $\theta$. 

Considering the metric of KS spacetime given in Eq.(\ref{eq:1}), we put the Hamilton-Jacobi action given in expression~(\ref{eq:4}) into Eq.(\ref{eq:3}). Due to the separation of variables, one can find easily the equations of motion of photons in the KS spacetime in the integrated and separated form~(see also \cite{Stuchlik14}),
\begin{eqnarray} 
\frac{dt}{d\lambda}&=&E f^{-1}(r) \ , \label{eq:5} \\
\frac{dr}{d\lambda}&=&\pm \sqrt{{R(r)}} \ , \label{eq:6} \\
\frac{d \theta}{d\lambda}&=& \pm \frac{\sqrt{\mathcal{Q}-\frac{L^2}{\sin^2 \theta}}}{r^2} \ , \label{eq:7} \\
\frac{d \phi}{d\lambda}&=& \frac{L}{r^2 \sin^2 \theta} \ , \label{eq:8}
\end{eqnarray}
where, $R(r)=\left[E^2-\mathcal{Q} f(r)/r^2\right]$, $\mathcal{Q}$ is the separation ``Carter'' constant~\cite{Carter68} having in the spherically symmetric spacetime direct meaning of square of the total angular momentum, and $f(r)$ is given by Eq.(\ref{eq:2}). We introduce two dimensionless impact parameters $\xi=L/E$ and $\eta=Q/E^2$ . To obtain the expressions of the impact parameters of the photon circular orbit (being the boundary of unstable circular orbits), we have to solve simultaneously equations $R(r)=0=dR(r)/dr$. We get that the dimensionless impact parameter $\eta$ corresponding to the photon circular orbit is determined by the relation 
\begin{equation} \label{eq:9}
\eta=r^2 f^{-1}(r) \ .
\end{equation}
\subsection{Circular geodesics}
We calculate the radius, energy and axial angular momentum of circular geodesics at the equatorial plane~($\theta=\pi/2$ and $p_\theta=0$) by solving the Hamiltonian-Jacobi equation~(\ref{eq:3}). Let us consider that the test particle has unit mass. In this case the Hamilton-Jacobi equation~(\ref{eq:3}) takes the form 
\begin{eqnarray} 
-\frac{E^2}{f(r)}+\frac{L^2}{f^2(r)}+f(r)\left(\frac{\partial S}{\partial r}\right)^2&=&-1 \nonumber \ ,  \\
\implies p_r^2=\left(\frac{\partial S}{\partial r}\right)^2&=&\mathcal{D} \ , \label{eq:10}
\end{eqnarray} 
where, $\mathcal{D}=f^{-2}(r) \left[E^2- f(r)(L^2/r^2+1)\right]$\ .
Using the condition giving the circular orbits, $p_r=0=\dot{p_r}$, we get from Eq.~(\ref{eq:10}) the expressions for the energy~($E$), and the axial angular momentum~($L$) in the form 
\begin{eqnarray} 
E^2=\frac{2f^2(r)}{2f(r)-f'(r)}  \ , \label{eq:11} \\
L^2=\frac{f'(r) r^3}{2f(r)-f'(r)} \ . \label{eq:12}
\end{eqnarray}
Photon circular orbit can be defined as the orbit where energy and angular momentum diverge. We can calculate the radius of photon circular orbit by solving the equation 
\begin{equation} \label{eq:13}
2f(r)-f'(r)=0 \ .
\end{equation}
\subsection{Radius of the black hole shadow}
\begin{figure}[t!]
\begin{center}
\includegraphics[width=.95\linewidth]{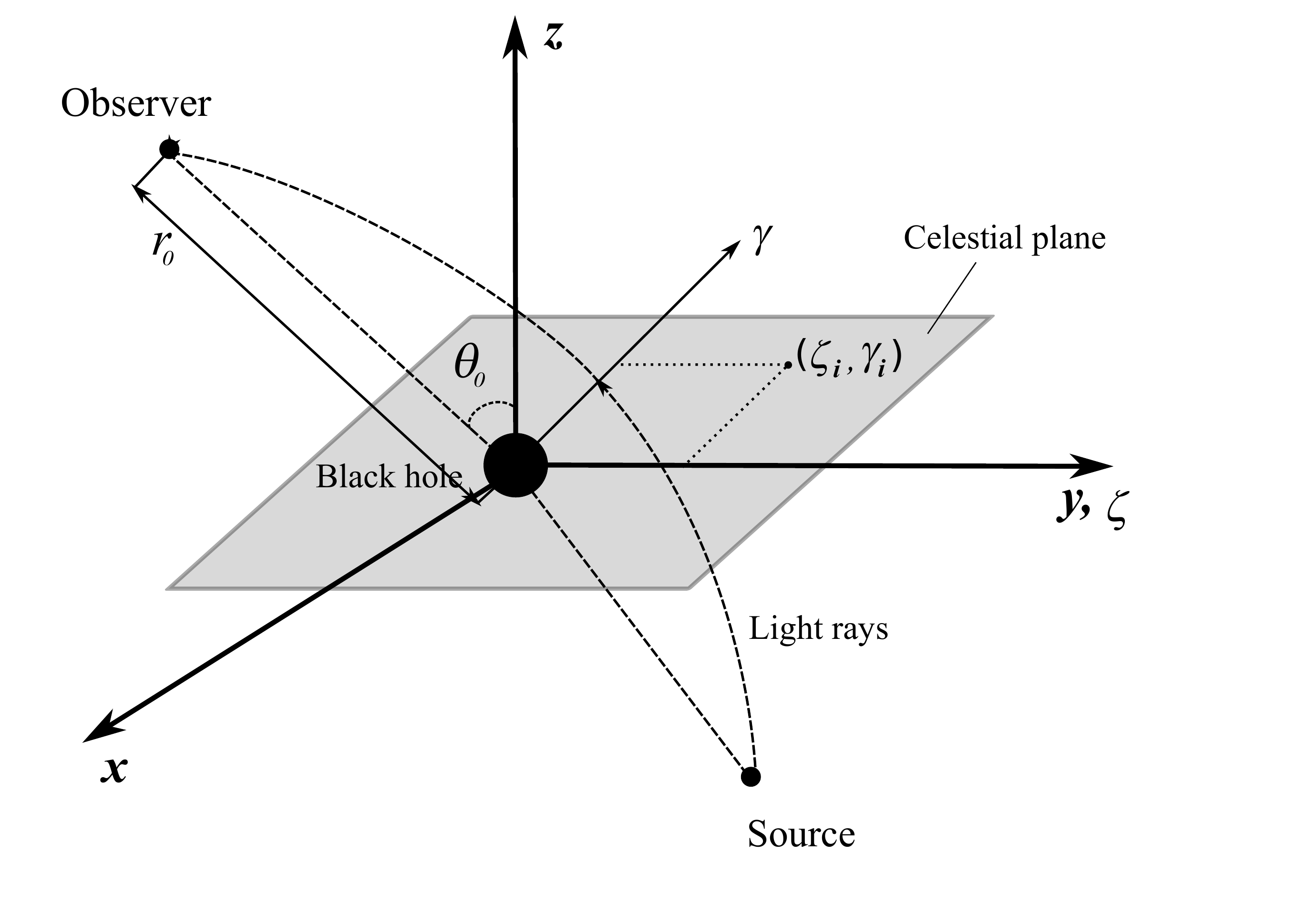} 
\end{center}
\caption{Demonstration of celestial coordinates. \label{celestial}}
\end{figure}
In order to analyse the apparent shape of black hole's shadow for distant observers, it is useful to introduce the celestial coordinates~(see \cite{hawking1973black,Barden1973ApJC,Vazquez04} for reference) defined by the relations 
\begin{eqnarray}
\zeta &=&\underset{r_0 \rightarrow \infty}{\lim}\left(-r_0^2 \sin \theta_0 \frac{d\phi}{dr}\right) \ , \label{eq:14}\\
\gamma &=&\underset{r_0 \rightarrow \infty}{\lim}\left(r_0^2 \frac{d\theta}{dr}\right) \ , \label{eq:15}
\end{eqnarray}
where $r_0$ is the distance between the observer and the black hole and $\theta_0$ is the inclination angle between the normal of observer's sky plane and observer-lens axis. We demonstrate definition of the celestial coordinates in Fig.~\ref{celestial}.
The celestial coordinates can be expressed in terms of the impact parameters determining the photon equations of motion~(\ref{eq:6},{\ref{eq:7},\ref{eq:8}) by the relations 
\begin{eqnarray}
\zeta &=&-\frac{\xi}{\sin \theta_0} \ , \label{eq:16} \\
\gamma &=&\sqrt{\eta-\frac{\xi^2}{\sin^2 \theta_0}} \ . \label{eq:17}
\end{eqnarray}
We have to plot `$\gamma$' vs `$\zeta$' in order to visualize the apparent shape of the image -- we can see from expressions~(\ref{eq:16}) and (\ref{eq:17}) that $\zeta^2+\gamma^2=\eta$, which implies that the apparent shape of the image is a circle of radius $\sqrt{\eta}$. After solving equation~(\ref{eq:13}), plugging the value of radius of photon circular orbit into expression of $\eta$ in (\ref{eq:9}) and calculating the square root of $\eta$, one can get the radius of apparent shape of black hole's shadow. The plot at the top left corner of Fig.~\ref{Shadow} depicts the apparent shape of the black hole's shadow in vacuum, given for typical values of the ``Ho\v rava'' parameter. 
\begin{figure*}
\begin{center}
\includegraphics[width=0.4\linewidth]{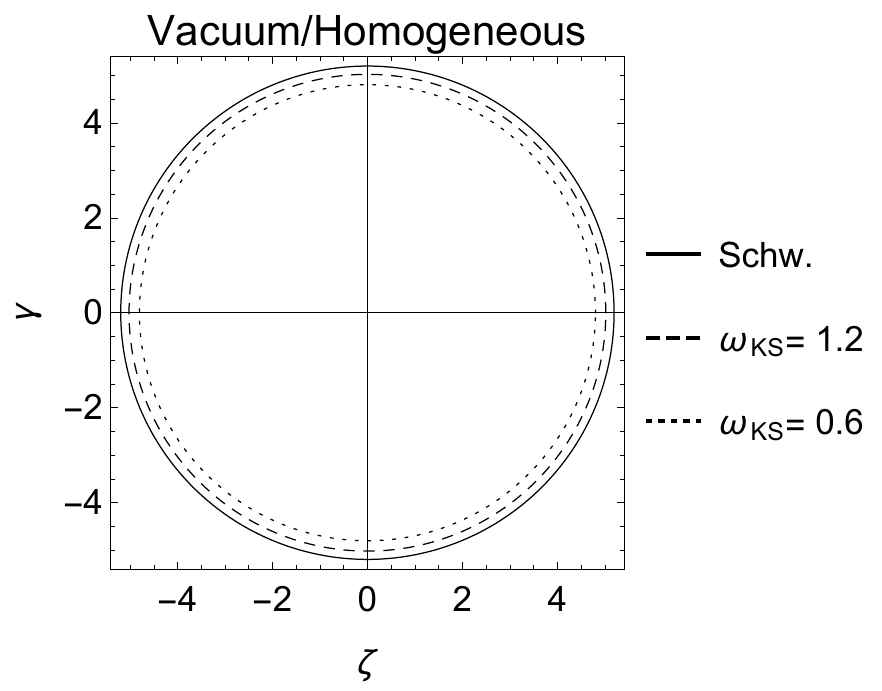} 
\includegraphics[width=0.4\linewidth]{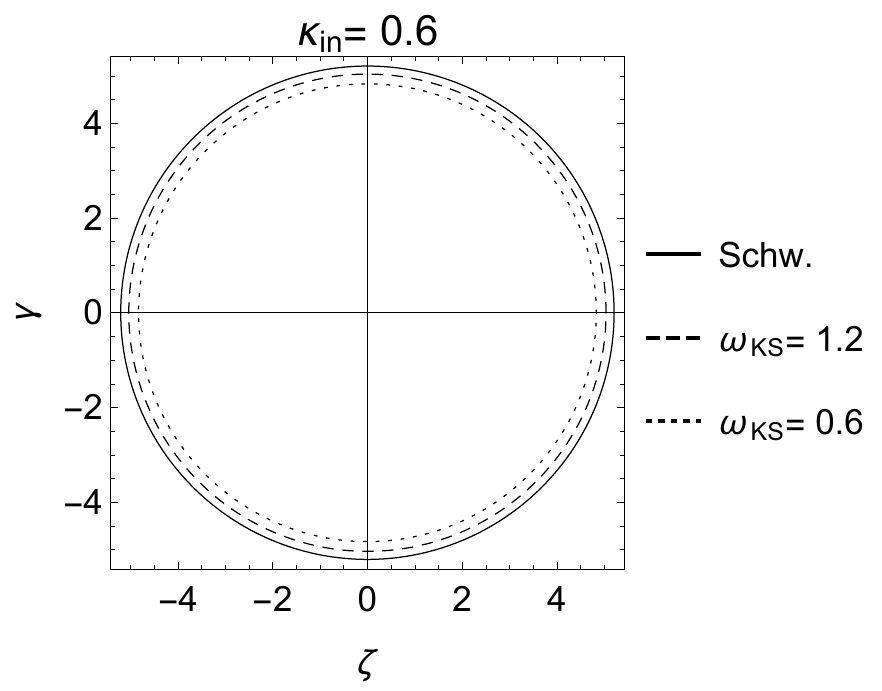}

\includegraphics[width=0.4\linewidth]{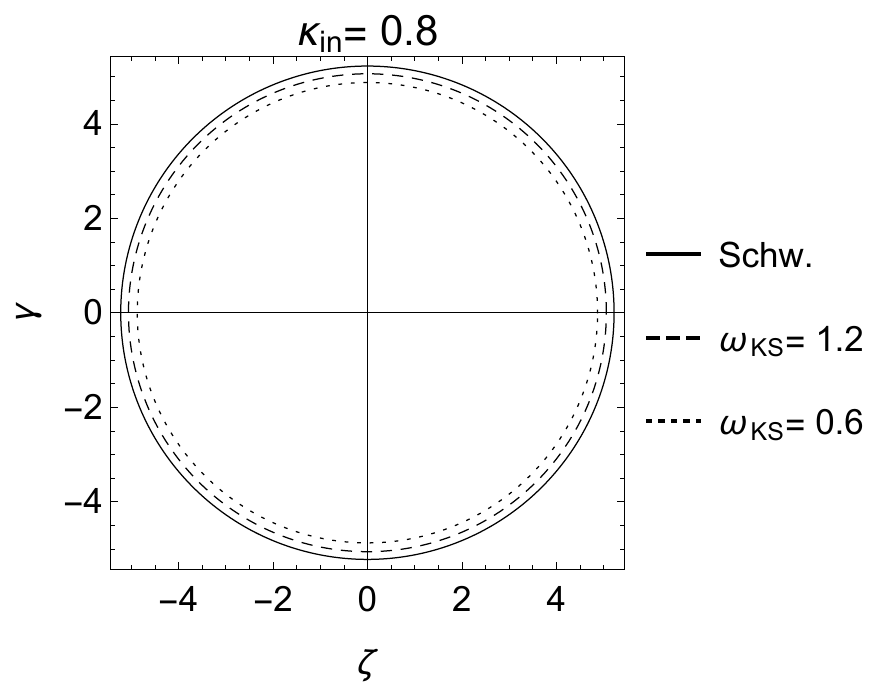}
\includegraphics[width=0.4\linewidth]{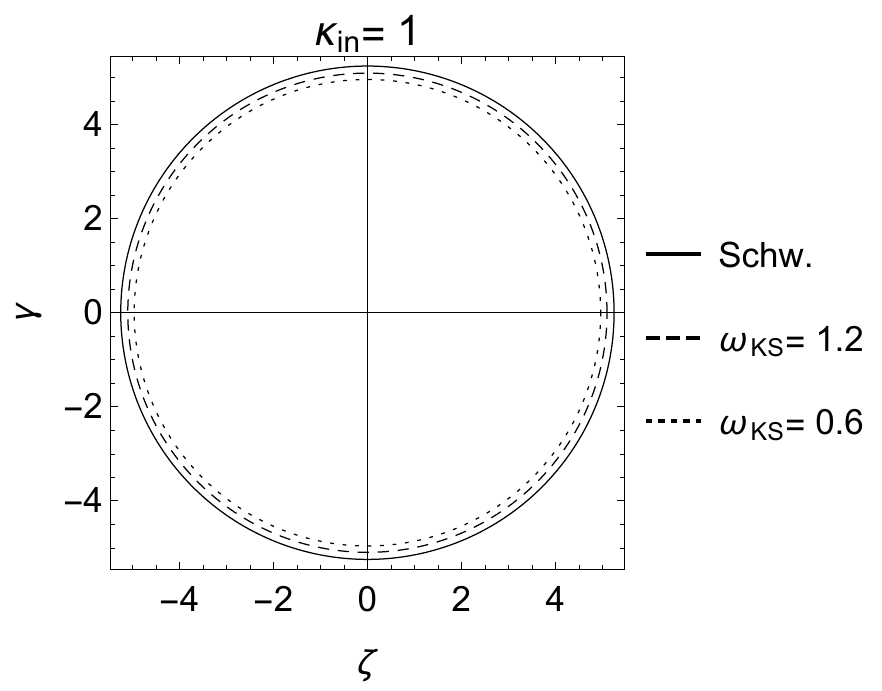}
 
\includegraphics[width=0.4\linewidth]{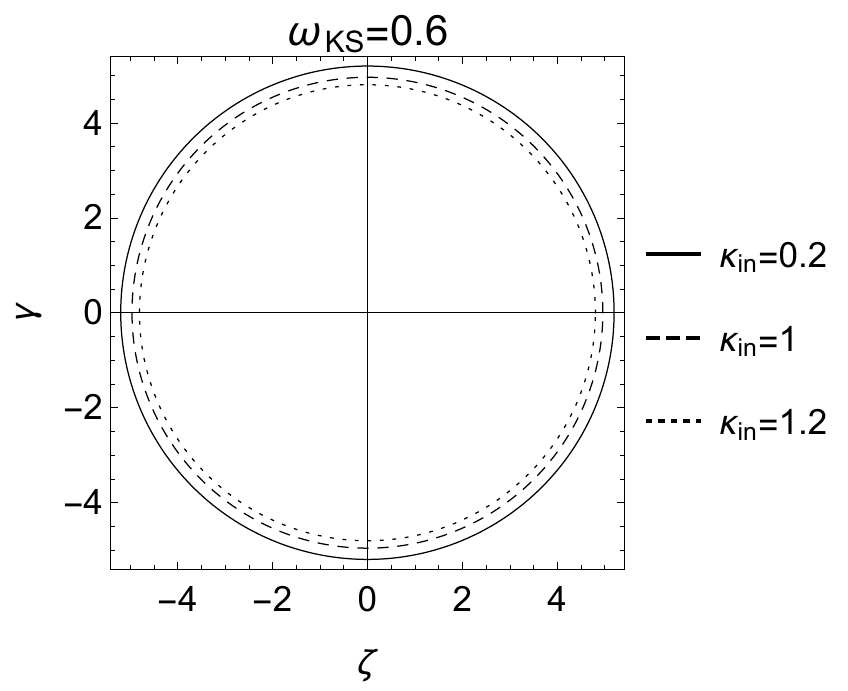}
\includegraphics[width=0.4\linewidth]{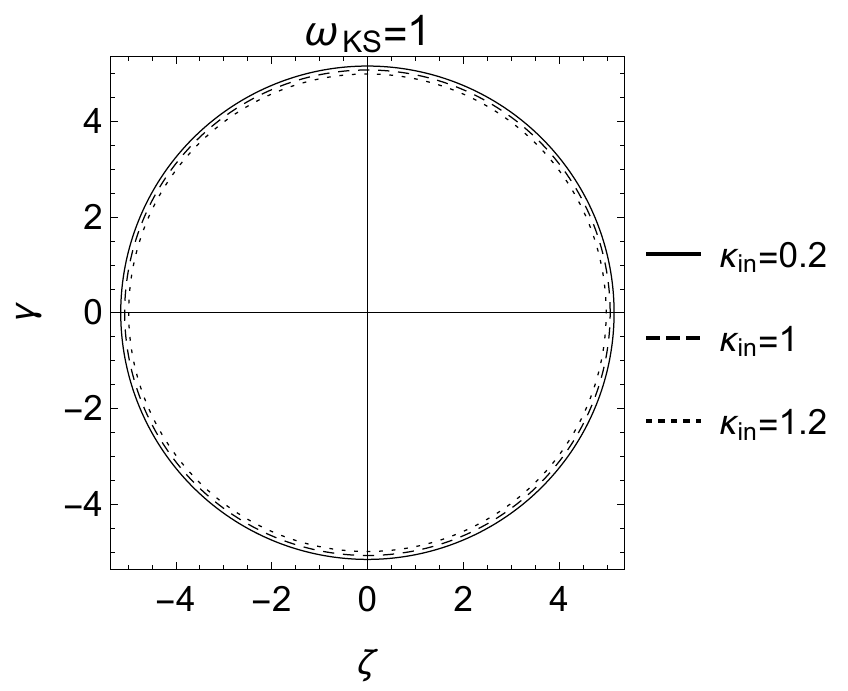}

\includegraphics[width=0.4\linewidth]{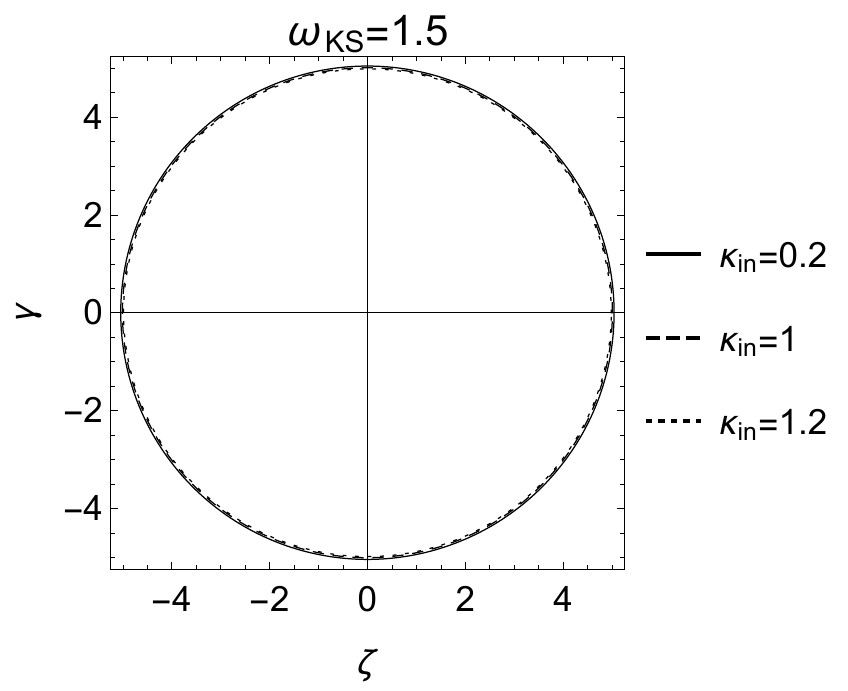}
\includegraphics[width=0.4\linewidth]{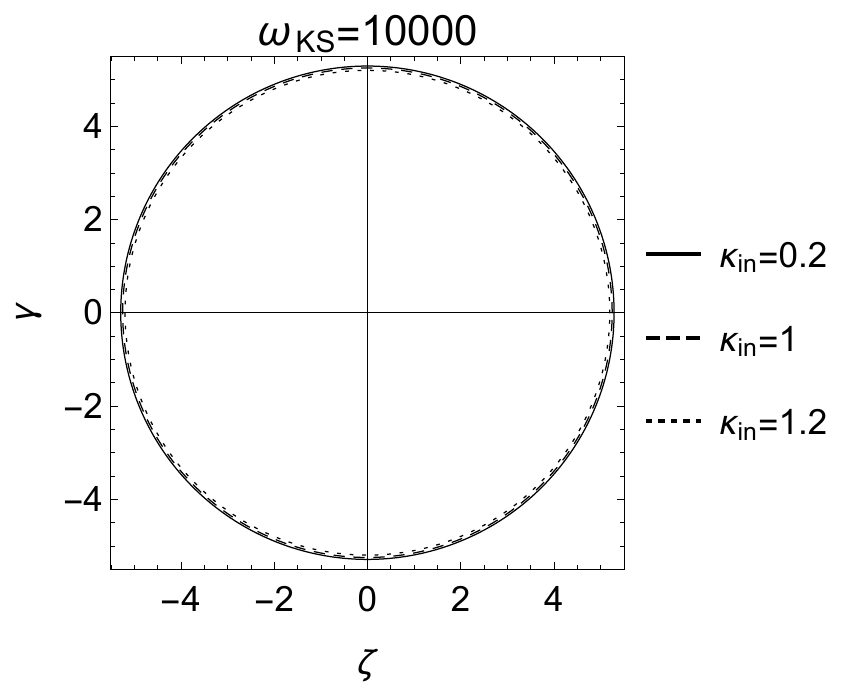} 
\end{center}
\caption{Shadows KS black holes. The top left corner image is for vacuum/homogeneous case and rest of the figures are plotted for various combinations of the ``Ho\v{r}ava'' parameter and the plasma parameter for inhomogeneous distribution of plasma. \label{Shadow}}
\end{figure*}
\section{Shadow of black hole in plasma \label{shadowplas} }
We consider a static distribution of plasma with refractive index $n$, which dependce on the photon frequency $\omega(x^i)$ is given by the relation 
\begin{equation} \label{eq:18}
n^2=1-\frac{\omega^2_e}{\omega^2(x^i)} \ , \hspace{0.25 cm} \omega^2_e=\frac{4\pi e^2 N}{m}=K_e N \ .
\end{equation}
Frequency of a photon~($\omega(x^i)$) depends on the spatial coordinates~($x^i$) as a result of the gravitational redshift. In  Eq.~(\ref{eq:18}), $N=N(x^i)$ is the electron number density in plasma, $m$ is the electron mass and $e$ is the electron charge.
\subsection{Photon motion in plasma}
The Hamilton-Jacobi equation in a static distribution of plasma background having refractive index `$n$' is given by the relation~\cite{Synge66,Rogers15,Bisnovatyi2010}
\begin{equation} \label{eq:19}
\frac{\partial S}{\partial \lambda}=\frac{1}{2} \left[g^{\mu \nu} \frac{\partial S}{\partial x^\mu} \frac{\partial S}{\partial x^\nu}-(n^2-1)\left(\frac{p_t}{\sqrt{-g_{tt}}}\right)^2 \right] \ .
\end{equation}
Again, considering the general form of metric~(\ref{eq:1}), we put the Jacobi action~($m=0$ in the case of photon) as given by Eq.~(\ref{eq:4}) into Hamilton-Jacobi equation~(\ref{eq:19}). Due to the possibility of separation of variables, one can easily arrive at the equations of photon motion in presence of plasma in the integrated and separated from
\begin{eqnarray} 
\frac{dt}{d\lambda}&=&n^2 E f^{-1}(r) \ , \label{eq:20} \\
\frac{dr}{d\lambda}&=&\pm \sqrt{{G(r)}} \ , \label{eq:21} \\
\frac{d \theta}{d\lambda}&=& \pm \frac{\sqrt{\mathcal{J}-\frac{L^2}{\sin^2 \theta}}}{r^2} \ , \label{eq:22} \\
\frac{d \phi}{d\lambda}&=& \frac{L}{r^2 \sin^2 \theta} \ , \label{eq:23}
\end{eqnarray}
where, $G(r)=\left[n^2 E^2-\mathcal{J} f(r)/r^2\right]$, $\mathcal{J}$ is the separation ``Carter'' constant~\cite{Carter68}, keeping the meaning of the total angular momentum in spherically symmetric backgrounds, and the lapse function $f(r)$ is given by Eq.(\ref{eq:1}).
We follow the same treatment as in the vacuum case, i.e., solving simultaneously equations $G(r)=0=dG(r)/dr$ to obtain the impact parameter $\eta$ of photon circular orbit given by the relation 
\begin{equation} \label{eq:24}
\eta=\frac{n^2 r^2}{f(r)} \ .
\end{equation}
\subsection{Radius of photon circular geodesic}
Following the same procedure as in the vacuum case, we use the Hamiltonian-Jacobi equation~(\ref{eq:19}) for photons and Eq.~(\ref{eq:21}) of their radial motion that gives the conditions of the circular photon motion $G(r)=0$, $dG(r)/dr=0$ implying the relations determining the radius of photon circular orbit 
\begin{equation} \label{eq:28}
2 n f(r)+2 n' r f(r)-n r f'(r)=0 \ .
\end{equation}
\subsection{Radius of the black hole shadow}
Using the equations of motion~(\ref{eq:21},{\ref{eq:22},\ref{eq:23}), we can get the relations giving the celestial coordinates~(\ref{eq:16},\ref{eq:17}) of the black hole shadow in terms of the photon impact parameters 
\begin{eqnarray}
\zeta &=&-\frac{\xi}{n \sin \theta_0} \ , \label{eq:29} \\
\gamma &=&\frac{1}{n} \sqrt{\eta-\frac{\xi^2}{\sin^2 \theta_0}} \ . \label{eq:30}
\end{eqnarray}
To visualize the apparent shape of the shadow image, we need to plot `$\gamma$' vs `$\zeta$'. In this case, we can see from Eqs~(\ref{eq:29}) and (\ref{eq:30}) that $\zeta^2+\gamma^2=\eta/n^2$, which implies that the apparent shape of the black hole shadow in plasma is a circle of radius $\sqrt{\eta}/n$. After solving Eq.~(\ref{eq:28}), plugging the value of radius of the photon circular orbit into expression of $\eta$ in (\ref{eq:23}) and calculating  $\sqrt{\eta}/n$, one can get the radius of apparent shape of black hole's shadow in the plasma. Fig.~\ref{Shadow} depicts the apparent shape of image of black hole's shadow in plasma in various cases discussed below. 
\subsection{Homogeneous distribution of plasma}
In this case the electron number density  is constant throughout the plasma distribution, i.e.
\begin{equation} \label{eq:31}
N=N_h=constant \ .
\end{equation}
So, from Eq.~(\ref{eq:18}) we get
\begin{equation} \label{eq:32}
n^2=1-\frac{K_eN_h}{\omega^2(x^i)}=1-\kappa_h \,
\end{equation}
where, $\kappa_h=K_eN_h/\omega^2(x^i)$. We call $\kappa_h$ as plasma parameter for homogeneous plasma distribution. Therefore, the refractive index `$n$ of the plasma is also constant and Eq.~(\ref{eq:28}) is reduced to Eq.~(\ref{eq:13}). This means that the radius of photon circular orbit remains the same as in the vacuum case. We also see that the radius of apparent image of black hole is the same as in the case of vacuum, i.e. $\sqrt{\eta}/n=\sqrt{r^2 f^{-1}(r)}$ \ . This case is similar as vacuum case.

\subsection{Inhomogeneous power-law distribution of plasma}
Here we consider that the number density of electrons in plasma is given by the relation
\begin{equation} \label{eq:33}
N=\frac{N_{in}r_0}{r} \,
\end{equation}
where $N_{in}$ is the number density of electrons at $r=r_0$ .
According to Eq.~(\ref{eq:18}), the refractive index of the medium~($n$) depends on position~($r$) as 
\begin{equation} \label{eq:34}
n=n(r)=1-\frac{K_e N_{in}r_0}{\omega^2(x^i) r}=1-\frac{\kappa_{in}}{r} \ ,
\end{equation} 
where a new plasma parameter $\kappa_{in}=K_eN_{in}r_0/\omega^2(x^i)$ is introduced. We call $\kappa_{in}$ as plasma parameter for inhomogeneous distribution of plasma. 

Using Eq.~(\ref{eq:34}), we can calculate the radius of photon circular orbit by solving Eq.~(\ref{eq:28}). Plugging the value of radius of photon circular orbit into Eq.~(\ref{eq:24}), we find the radius of the black hole shadow to be given given by $\sqrt{\eta}/n=\sqrt{r^2 f^{-1}(r)}$ \ . 

In Fig.~\ref{Shadow}, we plot the `$\gamma$' vs `$\zeta$' relations giving the black hole shadow for various combinations of the plasma coefficient $\kappa_{in}$, and the dimensionless ``Ho\v{r}ava'' parameter. We use 'Schw' abbreviation to refer the case of Schwarzschild limit. From top left corner plot and the second row plots, we see that the radius of the shadow is increasing with increasing plasma coefficient and it coincides with the Schwarzschild case for large values of the plasma coefficient. We also see from the figures of third and fourth row that radius of shadow is increasing with increasing dimensionless ``Ho\v rava'' parameter.

\section{Deflection of light near massive body surrounded by plasma in the weak field limit \label{photmotion}}
%
In Ref.~\cite{Kogan10} the authors introduced a special formalism for treating the gravitational lensing in the weak-field limit, for compact objects surrounded by plasma. In the present paper we use this formalism for calculation of the deflection angle of photons in weak gravitational field around KS black holes. 
 
We consider a static space-time with a metric 
\begin{equation} \label{eq:35}
ds^2=g_{\alpha \beta} dx^\alpha dx^\beta=g_{00}(dx^0)^2+g_{i j} dx^i dx^j \ .
\end{equation}

Assuming the weak-filed limit, we are allowed to write the metric as 
\begin{equation} \label{eq:36}
g_{\alpha \beta}=\eta_{\alpha \beta}+h_{\alpha \beta} \ ,
\end{equation}
here $\eta_{\alpha \beta}=\mathrm{diag}(-1,1,1,1)$ is the Minkowski metric of the flat spacetime, and  $h_{\alpha \beta}$ is a small perturbation satisfying conditions $ \left\lvert h_{\alpha \beta}\right\rvert\ll 1$ and $h_{\alpha \beta} \rightarrow 0$ for $\left\lvert x^i \right\rvert \rightarrow \infty$.
There is $\eta^{\alpha \beta}=\eta_{\alpha \beta}$ and $ h^{\alpha \beta}=h_{\alpha \beta} $.  

We can express the deflection angle as~\cite{Kogan10} 
\begin{equation} \label{eq:37}
\hat{\alpha}_i=\frac{1}{2} \int_{-\infty}^{+\infty} \left(h_{33,i}+\frac{\omega^2}{\omega^2-\omega^2_0} h_{00,i}-\frac{K_e}{\omega^2-\omega^2_0} N_{,i}  \right)dz \ ,
\end{equation}
where $i=1,2$ and $\omega_0=\omega_e(\infty)$. Using the definition of the deflection angle,
{\boldmath\begin{equation} \label{eq:38}
\hat{\mathbf{\alpha}}=\mathbf{e}(+\infty)-\mathbf{e}(-\infty) \ ,
\end{equation}}
we obtain the relation
\begin{equation} \label{eq:39}
\frac{de_i}{dz}=\frac{1}{2} \left(h_{33,i}+\frac{1}{n^2_0}h_{00,i}-\frac{1}{n^2_0 \omega^2} K_e N_{,i} \right)\ ,
\end{equation} 
for $i=1,2$ \ .
We can express the deflection angle given by Eq.(\ref{eq:37}) in terms of the impact parameter~$b$ as 
\begin{equation} \label{eq:40}
\hat{\alpha}_b=\frac{1}{2} \int_{-\infty}^{+\infty} \frac{b}{r} \left( \frac{dh_{33}}{dr}+\frac{\omega^2}{\omega^2-\omega^2_0} \frac{dh_{00}}{dr}- \frac{K_e}{\omega^2-\omega^2_0} \frac{dN}{dr}\right) dz \ ,
\end{equation}
where $r=\sqrt{b^2+z^2}$ \ .
For detailed exploration of this section, see ~\cite{Kogan10}.

\begin{figure}[t!]
\begin{center}
\includegraphics[width=0.9\linewidth]{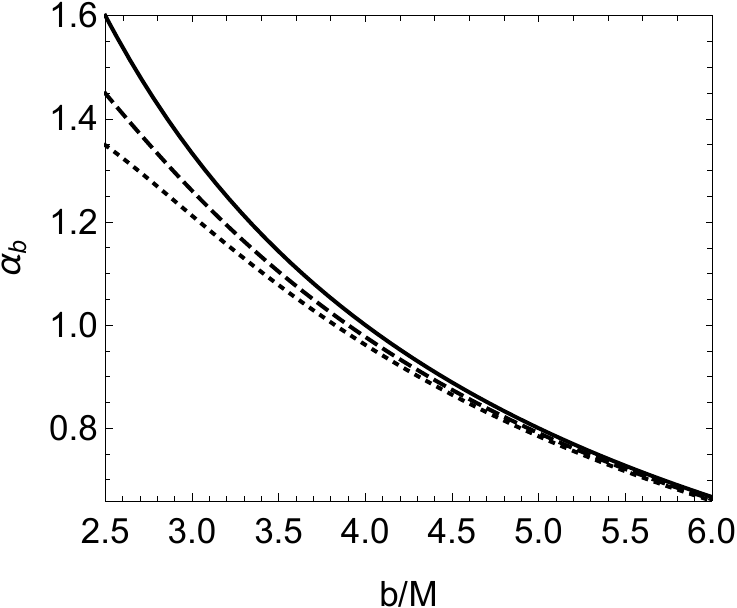}
\end{center}
\caption{The dependence of the deflection angle of photon on the dimensionless impact parameter~($b/M$) for various values of the dimensionless ``Ho\v rava'' parameter~($ \tilde{\omega}_{_{KS}}$) : 10000~(solid line), 1~(dashed line), 0.6~(dotted line).\label{defangl}}
\end{figure}
\begin{figure*}[t!]
\begin{center}
\includegraphics[width=0.3\linewidth]{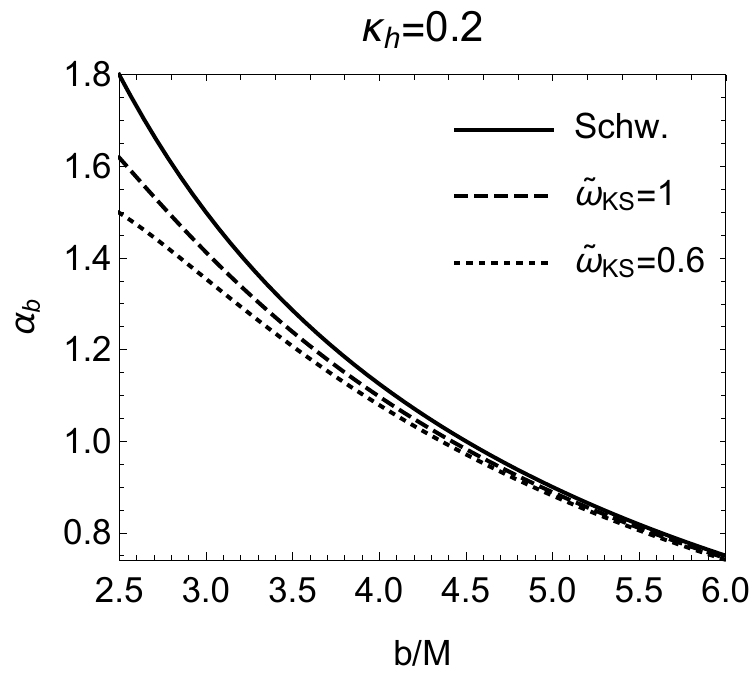} 
\includegraphics[width=0.3\linewidth]{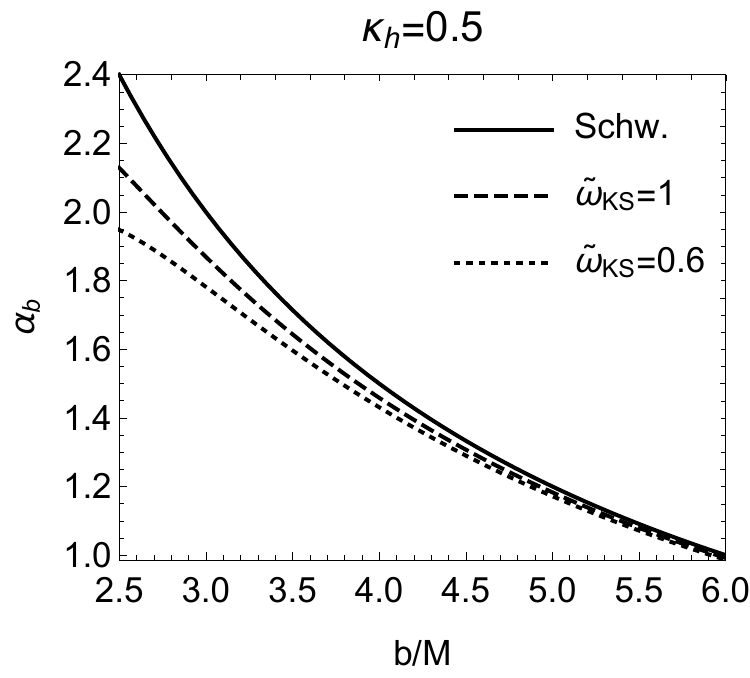}
\includegraphics[width=0.3\linewidth]{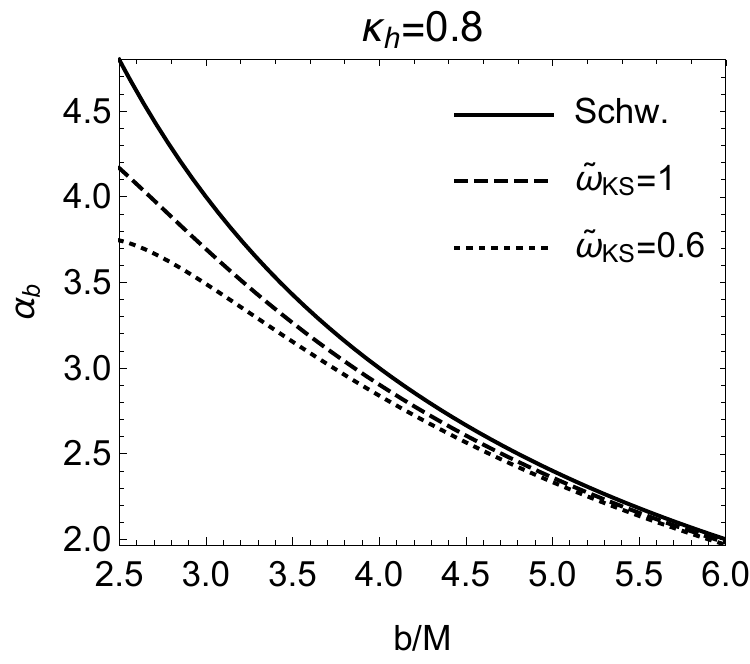}

\includegraphics[width=0.3\linewidth]{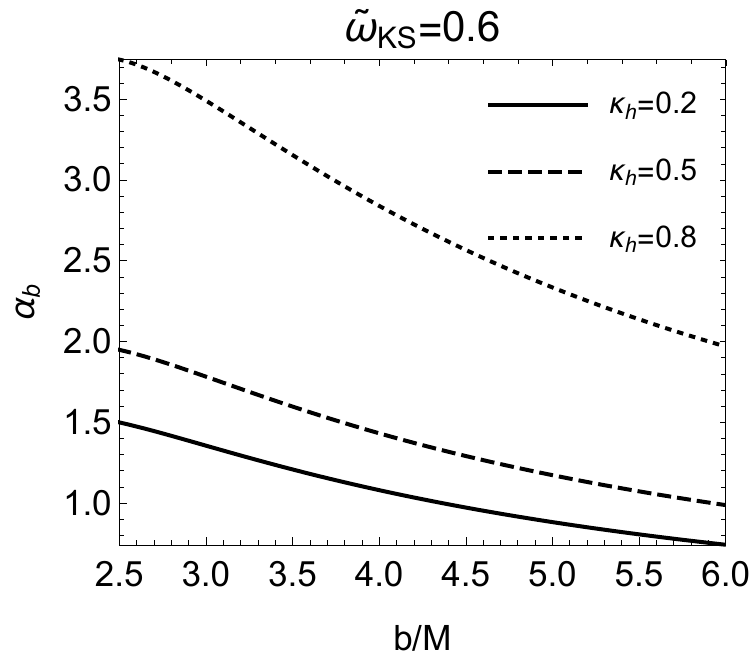} 
\includegraphics[width=0.3\linewidth]{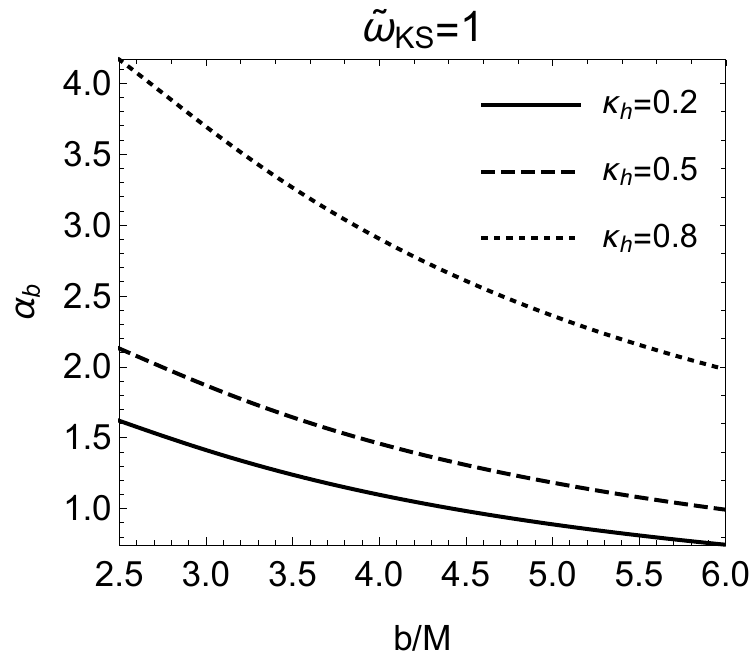}
\includegraphics[width=0.3\linewidth]{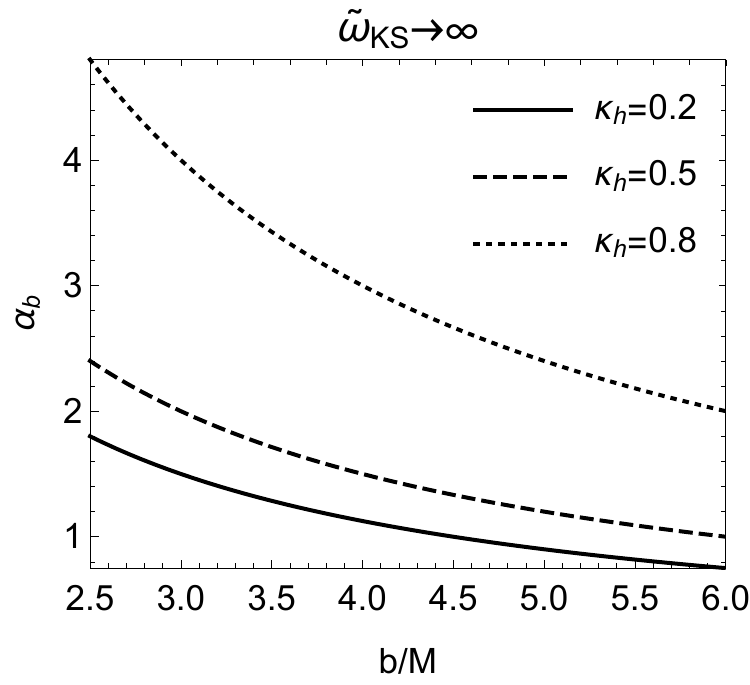} 

\includegraphics[width=0.3\linewidth]{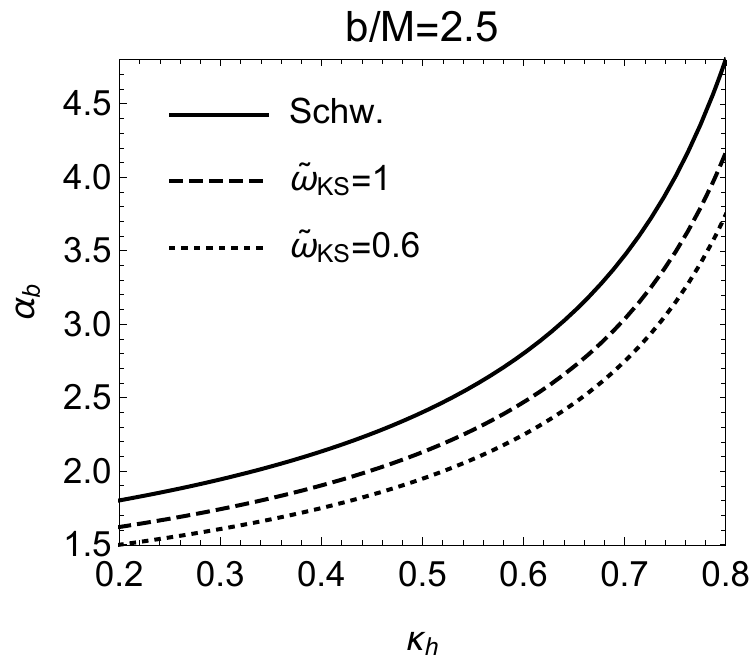}
\includegraphics[width=0.3\linewidth]{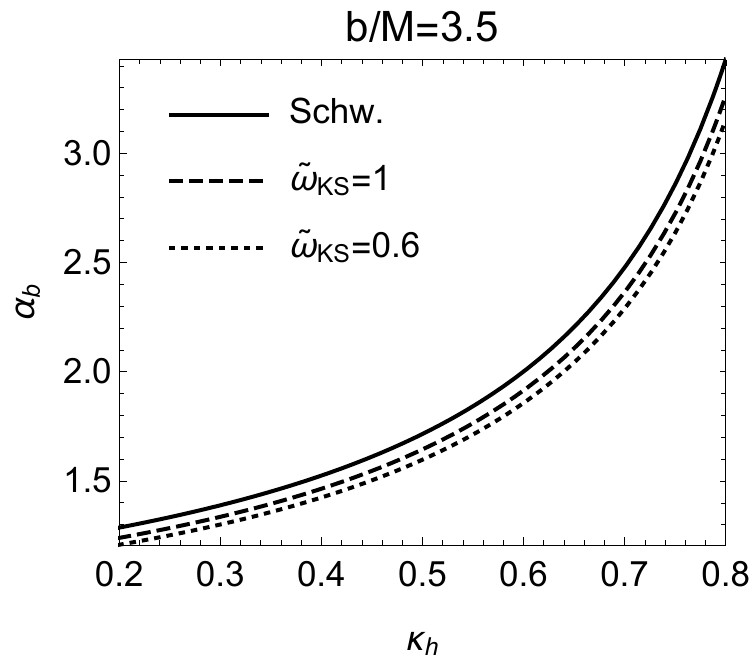} 
\includegraphics[width=0.3\linewidth]{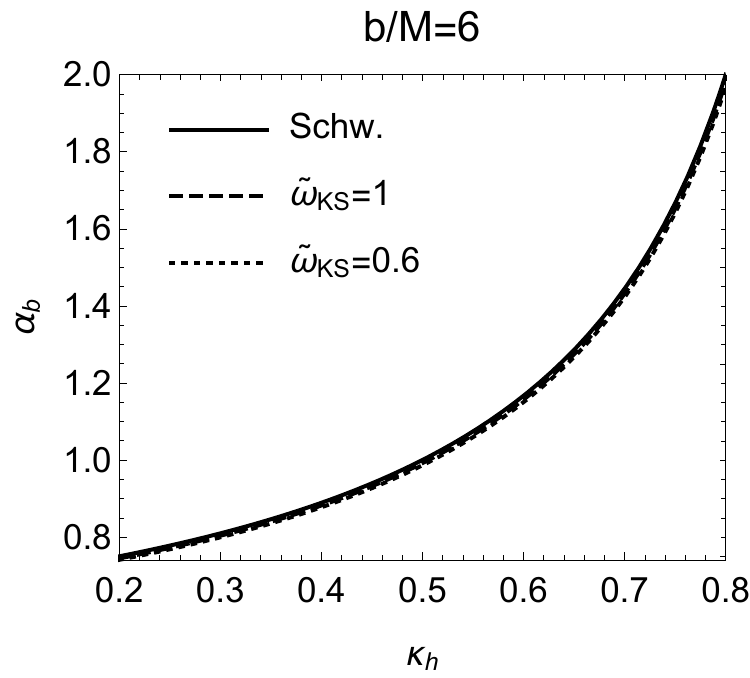}

\end{center}
\caption{\textbf{Homogeneous case:} In the first two rows, we show the dependence of the deflection angle of photon with dimensionless impact parameter~($b/M$) for various combinations of dimensionless ``Hora\v va'' parameter~($ \tilde{\omega}_{_{KS}}$) and plasma parameter.
In the third row, the dependence of the deflection angle of photon with plasma parameter for various combinations of dimensionless ``Hora\v va parameter"~($ \tilde{\omega}_{_{KS}}$) and dimensionless impact parameter~($b/M$). \label{uniformdefl}}
\end{figure*}
\begin{figure*}[t!]
\begin{center}
\includegraphics[width=0.3\linewidth]{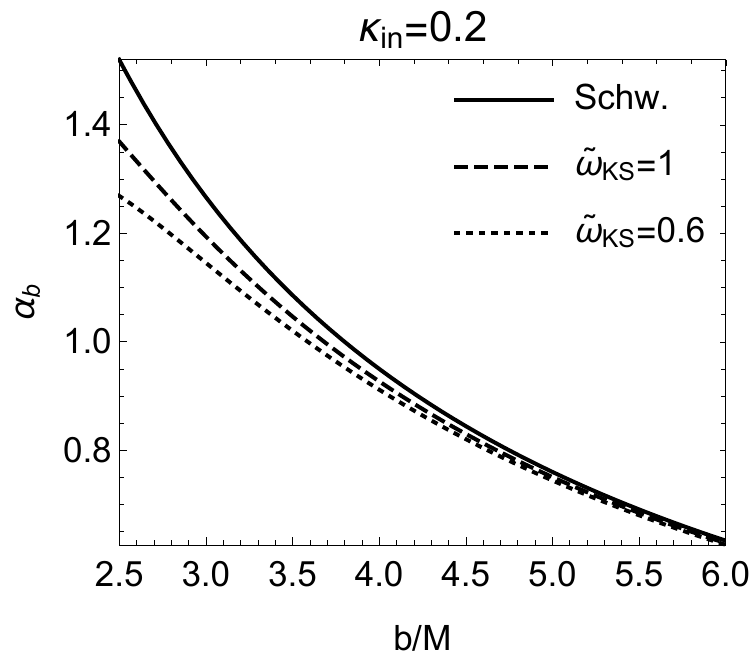} 
\includegraphics[width=0.3\linewidth]{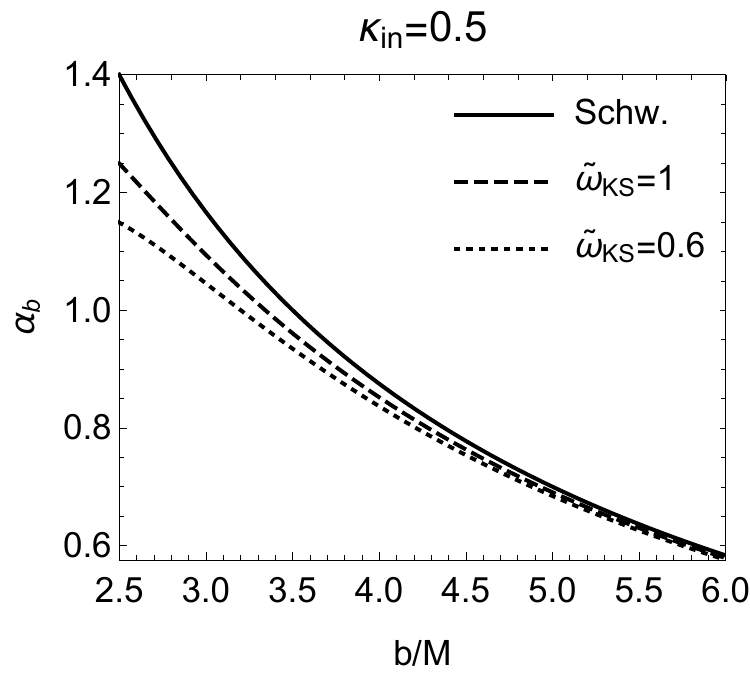}
\includegraphics[width=0.3\linewidth]{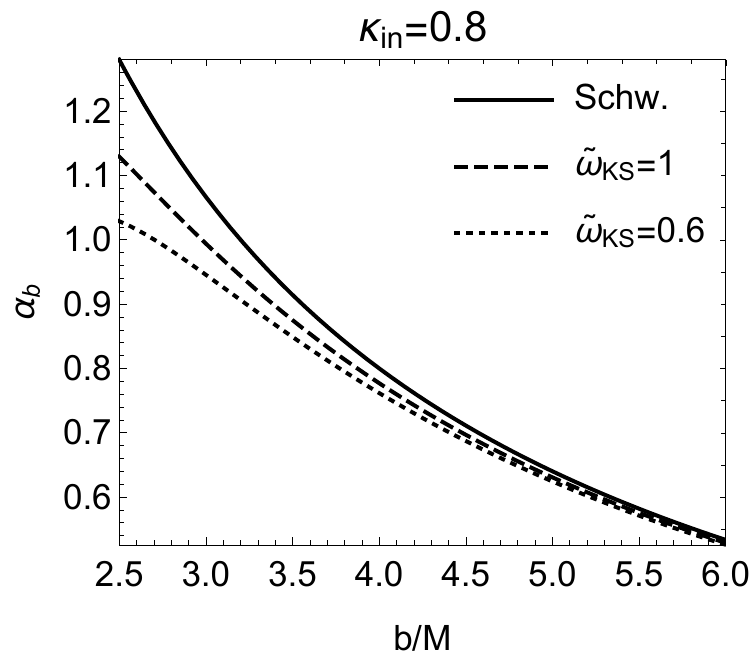}

\includegraphics[width=0.3\linewidth]{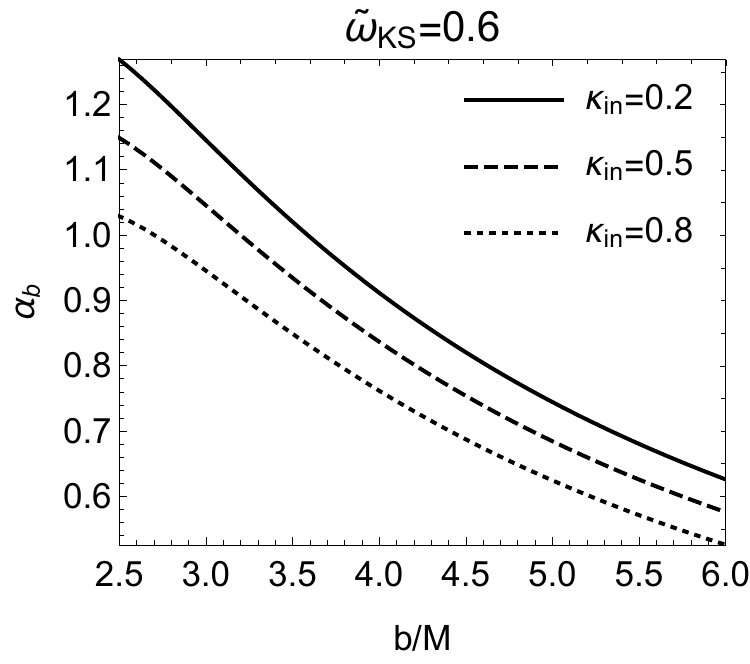} 
\includegraphics[width=0.3\linewidth]{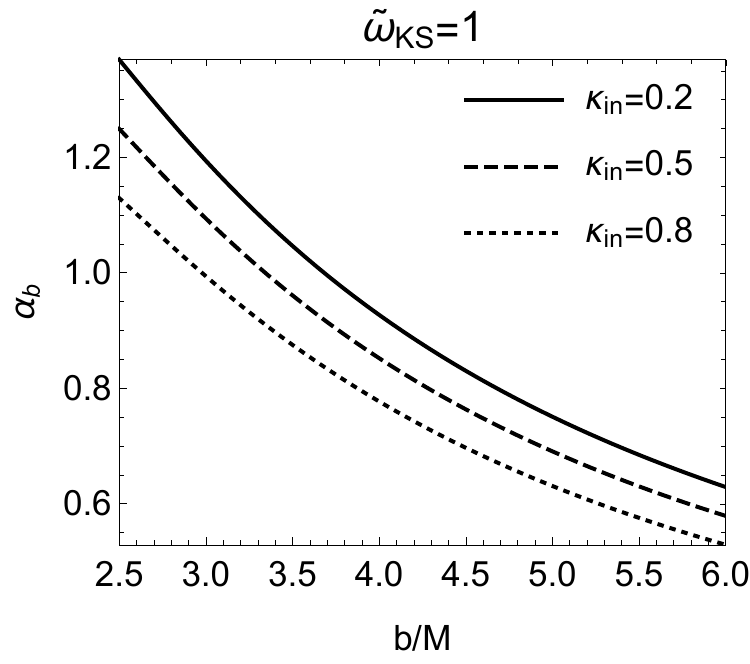}
\includegraphics[width=0.3\linewidth]{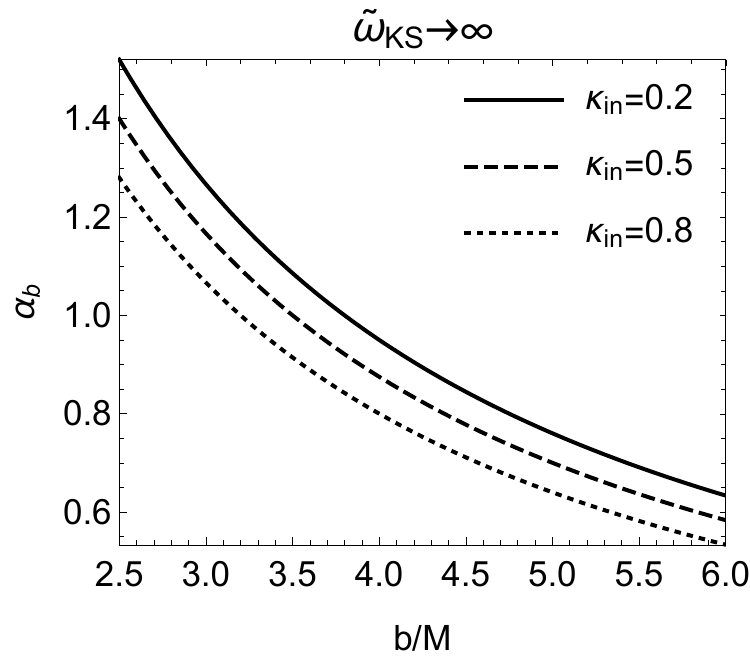} 

\includegraphics[width=0.3\linewidth]{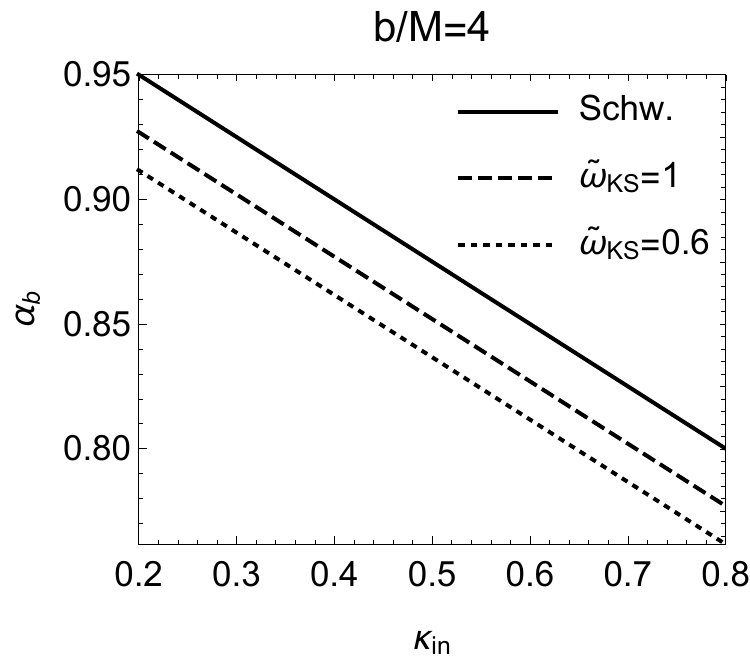}
\includegraphics[width=0.3\linewidth]{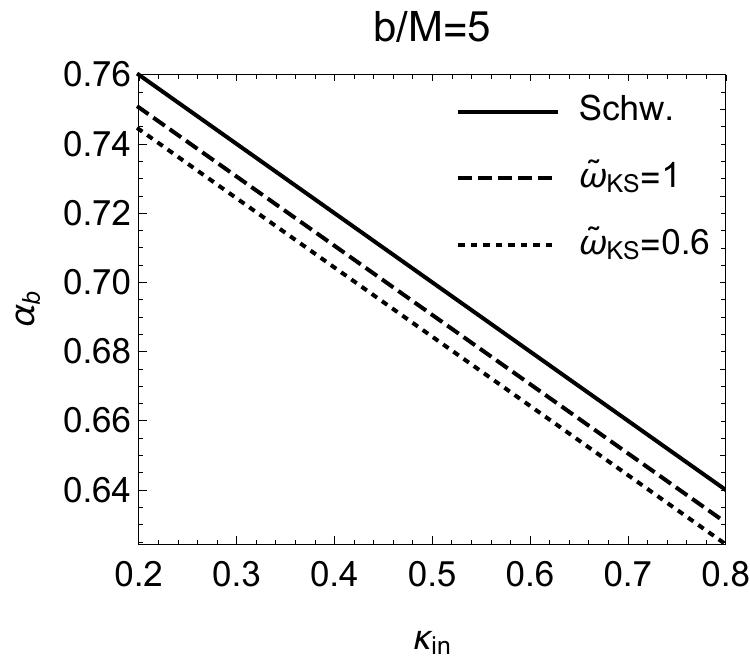} 
\includegraphics[width=0.3\linewidth]{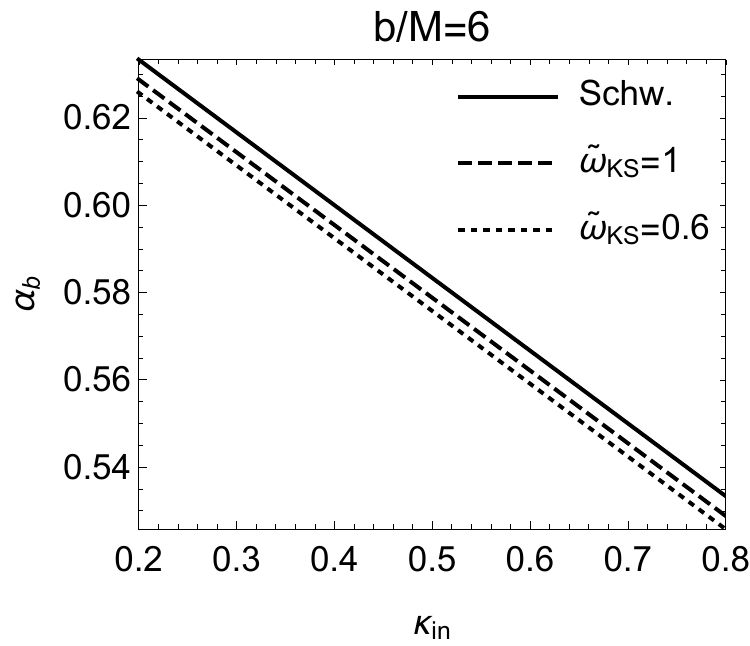}

\end{center}
\caption{\textbf{Inhomogeneous case:} In the first two rows, we show the dependence of the deflection angle of photon with dimensionless impact parameter~($b/M$) for various combinations of dimensionless ``Hora\v va'' parameter~($ \tilde{\omega}_{_{KS}}$) and plasma parameter.
In the third row, the dependence of the deflection angle of photon with plasma parameter for various combinations of dimensionless ``Hora\v va parameter"~($ \tilde{\omega}_{_{KS}}$) and dimensionless impact parameter~($b/M$). \label{powerdefl}}
\end{figure*}
Now, we calculate the deflection angle of photon by using the binomial approximation for the KS metric~(\ref{eq:1}). Considering $4M/\omega_{_{KS}}r^3 \ll 1$, and the binomial expansion 
\begin{equation}\label{eq:41}
(1+x)^n=1+nx+\frac{n(n-1)}{2!}x^2+\frac{n(n-1)(n-2)}{3!}x^3+... \ ,
\end{equation}
we can write 
\begin{eqnarray}
\left(1+\frac{4M}{\omega_{_{KS}} r^3}\right)^{\frac{1}{2}}&=&1+\frac{2M}{\omega_{_{KS}} r^3}-\frac{2M^2}{\omega_{_{KS}}^2 r^6}+\frac{4M^3}{\omega_{_{KS}}^3 r^9}-... \ ,  \nonumber \\
&& \left[\mathrm{if} \ , \frac{4M}{\omega_{_{KS}} r^3} \ll 1 \right].  \nonumber
\end{eqnarray}
Considering the binomial approximation and neglecting the higher order terms, the temporal and radial metric components can be written as
\begin{eqnarray} \label{eq:42}
-g_{tt}&=&f(r) \ , \nonumber \\
&=&1+r^2 \omega_{_{KS}} \left[\frac{2M^2}{\omega_{_{KS}}^2 r^6}-\frac{2M}{\omega_{_{KS}} r^3}-\frac{4M^3}{\omega_{_{KS}}^3 r^9}\right]  \ , \nonumber \\
&=&1-\frac{2M}{r}+\frac{2M^2}{\omega_{_{KS}} r^4}-\frac{4M^3}{\omega_{_{KS}}^2 r^7} \ ,
\end{eqnarray}
and
\begin{eqnarray} \label{eq:43}
g_{rr}&=&f(r)^{-1} \ , \nonumber \\
&=&\left[1+\left(\frac{2M^2}{\omega_{_{KS}} r^4}-\frac{2M}{r}-\frac{4M^3}{\omega_{_{KS}}^2 r^7} \right) \right]^{-1} \ , \nonumber \\
&=&1+\frac{2M}{r}-\frac{2M^2}{\omega_{_{KS}} r^4}+\frac{4M^3}{\omega_{_{KS}}^2 r^7} \ , \\
&& \left[\mathrm{if,} \quad \frac{2M^2}{\omega_{_{KS}} r^4}-\frac{2M}{r}-\frac{4M^3}{\omega_{_{KS}}^2 r^7}\ll 1 \right] \ . \nonumber
\end{eqnarray}
Considering the weak-field approximation~(\ref{eq:36}), the component of the metric tensor $h_{\alpha \beta}$ in the Cartesian coordinates have the following form: 
\begin{eqnarray} \label{eq:44}
h_{00}&=&\frac{2M}{r}-\frac{2M^2}{\omega_{_{KS}} r^4}+\frac{4M^3}{\omega_{_{KS}}^2 r^7} \ ,  \\ \label{eq:45}
h_{ik}&=&\left(\frac{2M}{r}-\frac{2M^2}{\omega_{_{KS}} r^4}+\frac{4M^3}{\omega_{_{KS}}^2 r^7} \right)s_i s_k \hspace{0.1cm} \ ,
\end{eqnarray}	
where $s_i$ correspond to the unit vectors along the coordinate axes: $s_1=x_1/r$ ,	$s_2=x_2/r$ , and $s_3=z/r=z/(z^2+b^2)^{1/2}=\cos \theta $. 
%

\subsection{Deflection angle in vacuum}
%

%
Now, we get the expression for the deflection angle of~photon in vacuum by inserting the electron plasma frequency ($\omega_0$) and the electron number density ($N(x^i)$) equal to zero in Eq.~(\ref{eq:40}). We then arrive to 
\begin{equation} \label{eq:46}
\alpha_{b}=\frac{1}{2} \int_{-\infty}^{+\infty} \frac{b}{r} \left(\frac{dh_{33}}{dr}+\frac{dh_{00}}{dr}\right)dz \ ,
\end{equation}
where $r=\sqrt{b^2+z^2}$ . Introducing the metric approximation and performing the above integral, we get
\begin{equation} \label{eq:47}
\alpha_{b}=-\frac{4M}{b}+ \frac{15 \pi}{8} \frac{M^2}{b^4 \omega_{_{KS}}}-\frac{512}{35} \frac{M^3}{b^7 \omega_{_{KS}}^2} \ .
\end{equation}
We can see that the result in expressed by Eq.~(\ref{eq:47}) is in agreement with the deflection angle in  the Schwarzschild geometry, i.e., $\alpha_b=-4M/b$ when $\omega_{_{KS}} \rightarrow \infty $. 

In Fig.~\ref{defangl}, we plot dependence of the photon deflection angle on the dimensionless impact parameter for typical values of the dimensionless ``Ho\v rava'' parameter. We can see that the ``Ho\v rava'' parameter demonstrate notable influence on photon deflection, if the impact parameter is low enough to allow motion close enough to the compact object. We see that the deflection angle coincides with the Schwarzschild case far from the compact object. 
%
\subsection{Deflection angle in homogeneous distribution of plasma}
%
Now we consider simple case when plasma distribution follows Eq.~(\ref{eq:31}). Using Eq. (\ref{eq:40}), we get
\begin{equation} \label{eq:48}
\alpha_{b}=\frac{1}{2} \int_{-\infty}^{+\infty} \frac{b}{r} \left(\frac{dh_{33}}{dr}+\frac{\omega^2}{\omega^2-\omega_0^2}\frac{dh_{00}}{dr} \right)dz \ ,
\end{equation}
Performing the above integral in the approximate form of the metric, we arrive at 
\begin{eqnarray} \label{eq:49}
\alpha_{b}&=&-\frac{2M}{b}\left(1+\frac{\omega^2}{\omega^2-\omega_0^2}\right)+\frac{3\pi M^2}{2 b^4 \omega_{_{KS}}}\left(\frac{1}{4}+\frac{\omega^2}{\omega^2-\omega_0^2}\right) \nonumber \\
 && -\frac{64 M^3}{5 b^7 \omega_{_{KS}}^2} \left(\frac{2}{15}+\frac{ \omega^2}{\omega^2-\omega_0^2}\right) \ .
\end{eqnarray}
Using Eq.~(\ref{eq:32}), we can write the above expression as
\begin{eqnarray} \label{eq:50}
\alpha_{b}&=&-\frac{2M}{b}\left(1+\frac{1}{1-\kappa_h}\right)+\frac{3\pi M^2}{2 b^4 \omega_{_{KS}}}\left(\frac{1}{4}+\frac{1}{1-\kappa_h}\right) \nonumber \\
 && -\frac{64 M^3}{5 b^7 \omega_{_{KS}}^2} \left(\frac{2}{15}+\frac{1}{1-\kappa_h}\right) \ ,
\end{eqnarray}
here we are neglecting the redshift of photon.

In the first and second rows of Fig. \ref{uniformdefl}, we demonstrate that the deflection angle is increasing as a result of increasing plasma parameter. We can see that this phenomenon is significant near the KS compact object, and the deflection is increasing with increasing ``Ho\v rava'' parameter.
In the third row of Fig. \ref{uniformdefl}, we demonstrate that plasma parameter has significant influence near the compact object; increasing of the plasma parameter implies increasing of the deflection angle. The effect of  the plasma parameter becomes weaker far from the KS compact object. So, if the plasma density near KS compact object increases, the deflection angle increases.

\subsection{Deflection angle in inhomogeneous distribution of plasma($N=N_0/r$)}
%

%
We assume that the number density of electrons in plasma varies with position due to Eq.~(\ref{eq:33}). Using Eq.~(\ref{eq:40}) with the metric coefficient in the approximate form, and performing the integral, we get the deflection angle 
\begin{eqnarray} \label{eq:51}
\alpha_{b}&=&-\frac{2M}{b}\left(1+\frac{\omega^2}{\omega^2-\omega_0^2}\right)+\frac{3\pi M^2}{2 b^4 \omega_{_{KS}}}\left(\frac{1}{4}+\frac{\omega^2}{\omega^2-\omega_0^2}\right) \nonumber \\
 && -\frac{64 M^3}{5 b^7 \omega_{_{KS}}^2} \left(\frac{2}{15}+\frac{ \omega^2}{\omega^2-\omega_0^2}\right)+ \frac{N_{in} K_er_0}{b(\omega^2-\omega_0^2)} \ .
\end{eqnarray}
Taking into account $\omega_0=\omega_e(\infty)=0$, and by using~(\ref{eq:34}), we rewrite the above expression as 
\begin{eqnarray} \label{eq:52}
\alpha_{b}=-\frac{4M}{b}+\frac{15\pi M^2}{8 b^4 \omega_{_{KS}}} -\frac{1088 M^3}{75 b^7 \omega_{_{KS}}^2} + \frac{\kappa_{in}}{b} \ .
\end{eqnarray}

In Fig.~\ref{powerdefl}, we show how the deflection angle is influenced by the plasma distribution near the compact object. In this case, deflection angle is significantly lower than in the case of homogeneous plasma distribution. Deflection angle decreases monotonically with increasing plasma parameter. This behaviour is opposite to the homogeneous case.
%

%
\section{The magnification of the source image \label{magn} }
%
In this section, we consider magnification of the brightness of the source image due to the weak gravitational lensing. We start from well known lens equation given in~\cite{Narayan96} 
\begin{equation} \label{eq:53}
\theta D_s=\beta D_s + \alpha_b D_{ls} \ ,
\end{equation}
where $\beta$ and $\theta$ are the source angle and the image angle, respectively, being related to the observer-lens axis, $D_s$ is the distance between the source and the observer, and $D_{ls}$ is the distance between the lens and the source. Manipulating the lens equation~(\ref{eq:53}), we obtain 
\begin{equation} \label{eq:54}
\beta=\theta- \frac{D_{ls}}{D_s} \frac{F(\theta)}{D_l} \frac{1}{\theta} \ ,
\end{equation}
where $F(\theta)=\left\lvert \alpha_b \right\rvert b=\left\lvert \alpha_b(\theta) \right\rvert D_l \theta$ , $D_l$ is the distance between the lens and the observer.
\begin{figure}[t!]
\begin{center}
\includegraphics[width=0.9\linewidth]{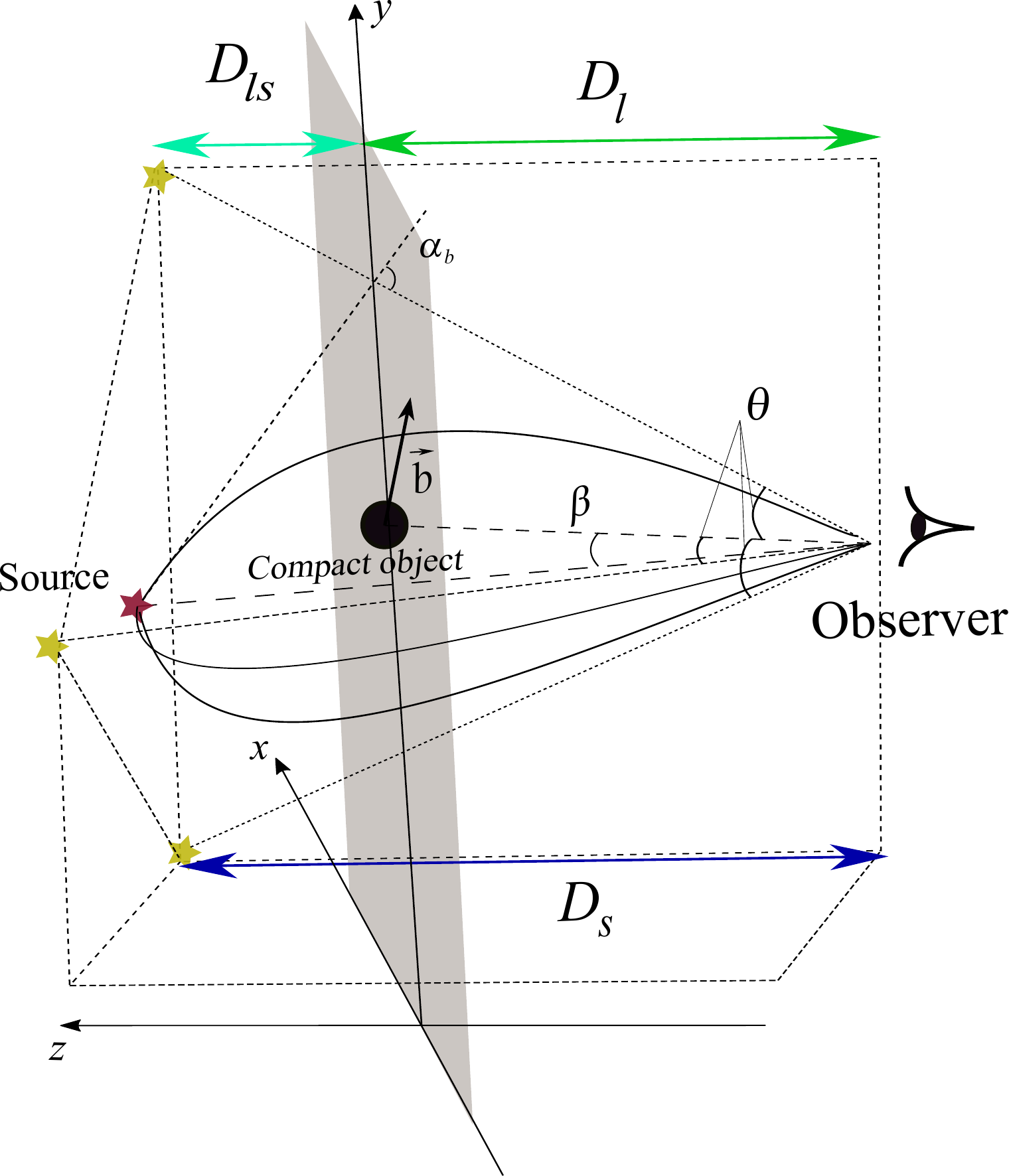}
\end{center}
\caption{Schematic plot of Source-Observer-lens system.\label{lenseq}}
\end{figure}
\begin{figure*}[t!]
\begin{center}
\includegraphics[width=0.4\linewidth]{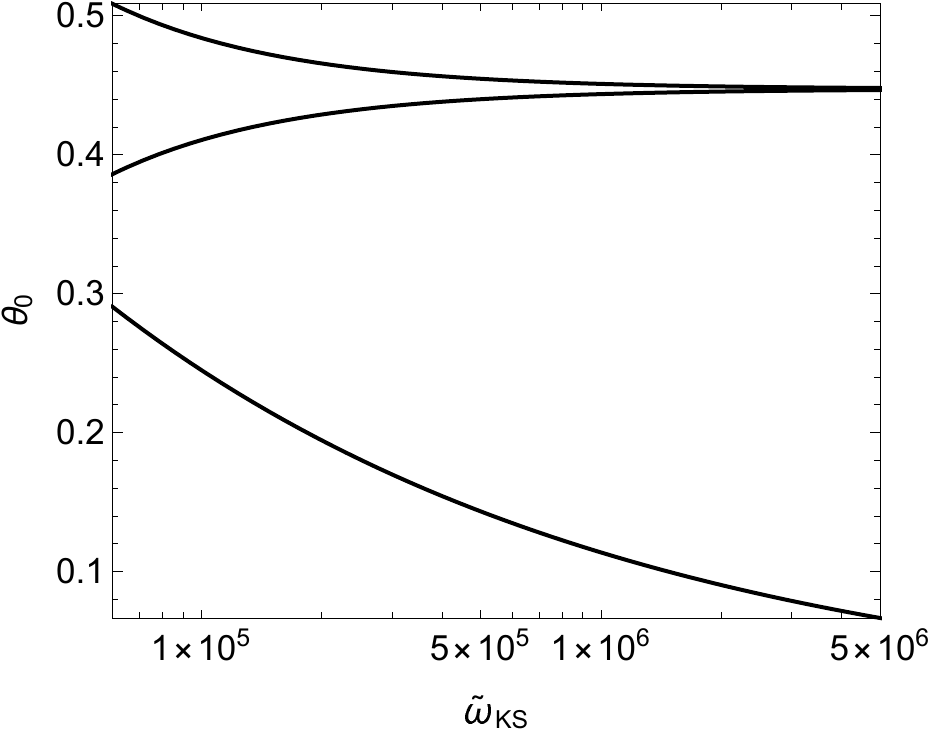} \hspace{0.25 cm}
\includegraphics[width=0.4\linewidth]{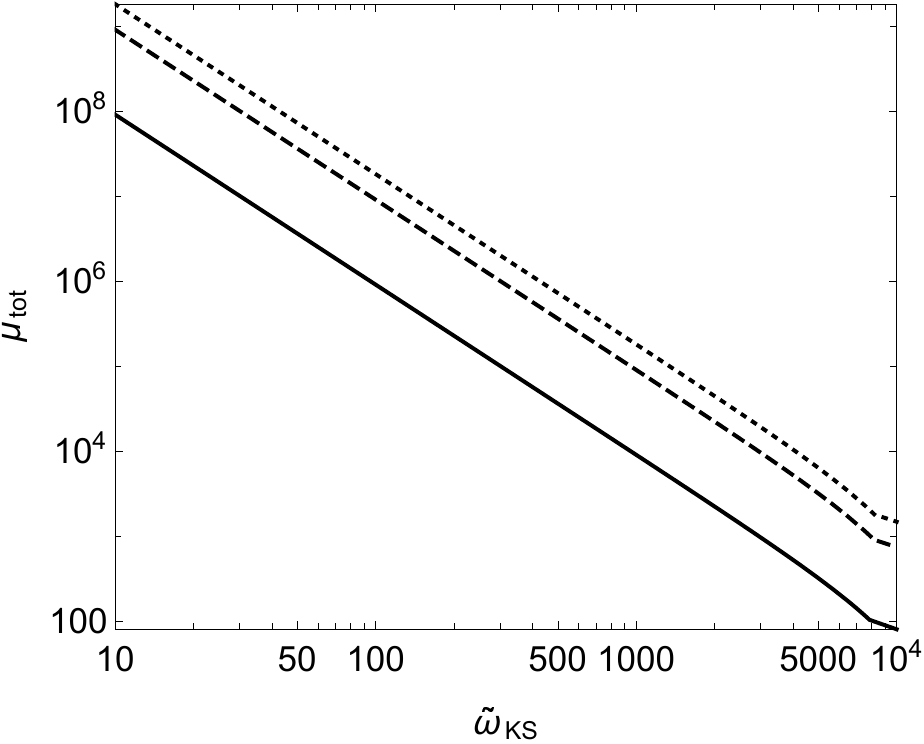}
\end{center}
\caption{\textbf{Vacuum case:} On the left panel, we plot dependence of three Einstein angles~($\theta_0$) with dimensionless ``Hora\v va parameter"~($ \omega_{_{KS}}$).
On the right panel, we plot the total magnification of image brightness with dimensionless ``Hora\v va parameter"~($ \omega_{_{KS}}$) for different values of angle of source from observer-lens axis($\beta$): 0.01~(solid line), 0.001~(dashed line), 0.0005~(dotted line). Both plots are considering $M/D_{l} = 10$ and $D_{ls}/D_{s} = 1/200 $ . \label{Einsangltotmag}}
\end{figure*}
\begin{figure*}[t!]
\begin{center}
\includegraphics[width=0.4\linewidth]{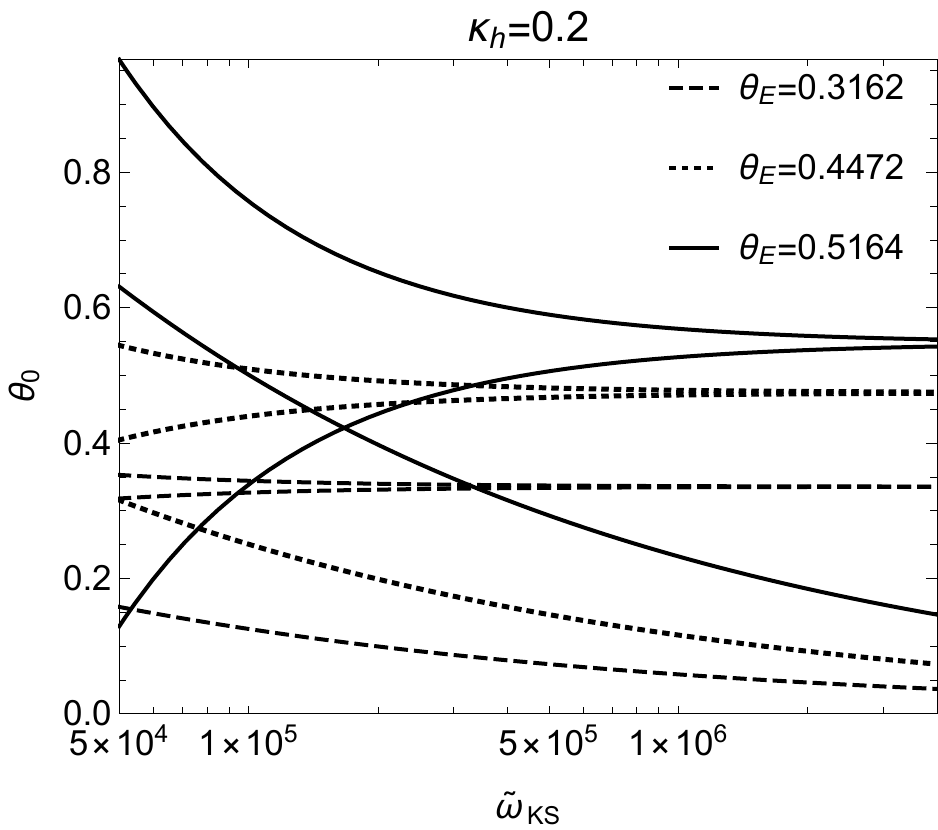} \hspace{0.7 cm}
\includegraphics[width=0.4\linewidth]{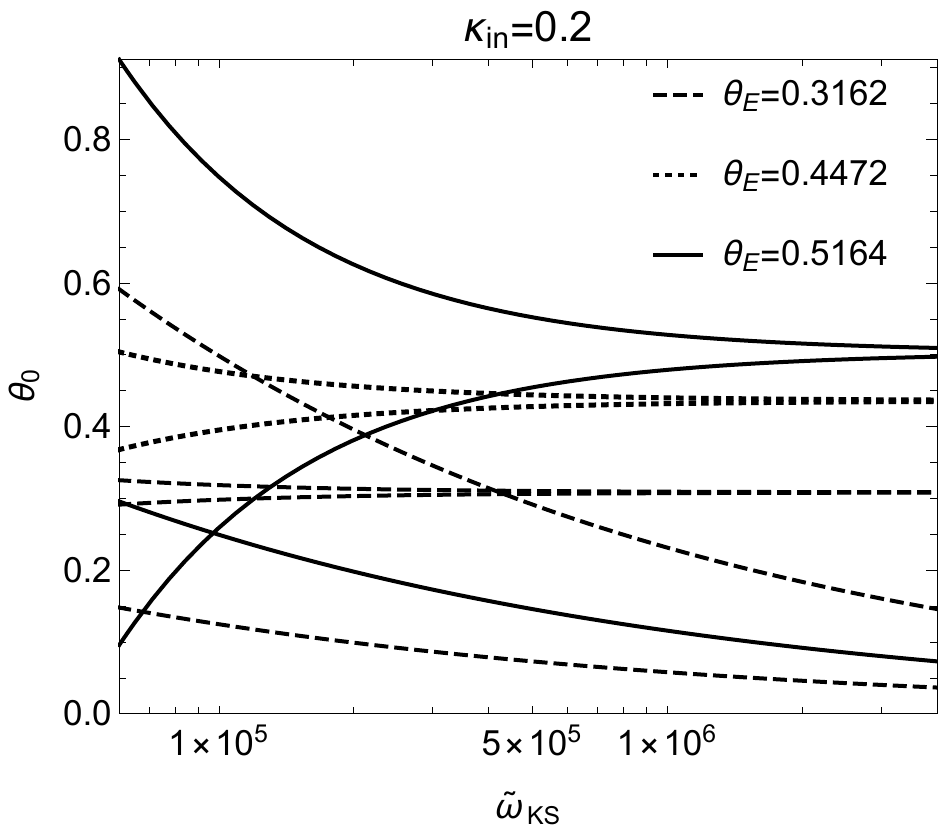}
\end{center}
\caption{On the left panel, we show the dependence of Einstein angle~($\theta_0$) for homogeneous distribution of plasma  with dimensionless ``Hora\v va parameter"~($ \omega_{_{KS}}$) for different values of $\theta_E$: 0.5164~(solid line), 0.4472~(dashed line), 0.3162~(dotted line).\\
On the right panel, we show the dependence of Einstein angle~($\theta_0$) for inhomogeneous distribution of plasma  with dimensionless ``Hora\v va parameter"~($ \omega_{_{KS}}$) for different values of $\theta_E$: 0.5164~(solid line), 0.4472~(dashed line), 0.3162~(dotted line). \label{unipwr}}
\end{figure*}
In Fig.~\ref{lenseq}, we present a schematic plot of the source-observer-lens system. Positions of images~($\theta_d$) formed due to the lensing can be found by solving Eq.~(\ref{eq:54}), where $d$ is the number of the image. Radius of the Einstein ring is defined by $R_0=D_l \theta _0$,  where $\theta_0$ is the solution of Eq.~(\ref{eq:54}) when the source is located on the line of sight i.e. $\beta=0$, and $\theta_0$ is defined as the Einstein angle. 

The magnification of the image brightness is defined by the formula
\begin{equation} \label{eq:55}
\mu_{tot}=\frac{I_{tot}}{I}=\sum_d \left\lvert \left(\frac{\theta_d}{\beta}\right) \left(\frac{d\theta_d}{d\beta}\right) \right\rvert \ , \hspace{0.2 cm} d=1,2,....,f,
\end{equation}
where $f$ is the number of images, $I_{tot}$ is the total brightness of images, $I$ is the brightness of the source and $d$ is the index of a concrete image.

\subsection{Image magnification in vacuum}
%
Using Eq.~(\ref{eq:47})~ and neglecting terms involving~$\omega^2$, Eq.~(\ref{eq:54}) takes the form
\begin{eqnarray} \label{eq:56}
\beta &=&\theta-\frac{4M}{D_l} \frac{D_{ls}}{D_s} \left(\frac{1}{\theta}-\frac{15 \pi}{32} \frac{M}{D_l^3 \omega_{_{KS}}} \frac{1}{\theta^3} \right) \ , \nonumber \\
&=& \theta-  \left(\frac{4M}{D_l} \frac{D_{ls}}{D_s} \right)\frac{1}{\theta} \nonumber \\ 
&&+\left(\frac{15 \pi}{32} \frac{M}{D_l^3 \omega_{_{KS}}} \right) \left(\frac{4M}{D_l} \frac{D_{ls}}{D_s} \right)\frac{1}{\theta^3} \ .
\end{eqnarray}
Introducing parameters,
\begin{eqnarray} 
\theta_E^2 &=&\frac{4M}{D_l} \frac{D_{ls}}{D_s} \ , \label{eq:57}\\
\theta_F &=&\left(\frac{15 \pi}{32} \frac{M}{D_l^3 \omega_{_{KS}}} \right) \left(\frac{4M}{D_l} \frac{D_{ls}}{D_s} \right) \ , \nonumber \\
&=&\left(\frac{15 \pi}{32} \frac{M}{D_l^3 \omega_{_{KS}}} \right) \theta_E^2 \label{eq:58}\ ,
\end{eqnarray}
Eq.~(\ref{eq:56}) takes the form
\begin{equation} \label{eq:59}
\theta^5-\beta \theta^4-\theta_E^2 \theta^3+\theta_F=0 \ .
\end{equation}
The above equation has three real roots depicting positions of the images. The solutions are given by
\begin{eqnarray}
\theta_1 &=&\frac{1}{2} \left[\beta+\mathcal{B}-\frac{16\theta_F}{\mathcal{B} (\beta+\mathcal{B})^3 }  \right] \label{eq:60} \ , \\
\theta_2 &=&\frac{1}{2} \left[\beta-\mathcal{B}+\frac{16\theta_F}{\mathcal{B} (\beta-\mathcal{B})^3 }  \right] \label{eq:61} \ , \\
\theta_3 &=& \left(\frac{\theta_F}{\theta_E^2}\right)^{\frac{1}{3}} \label{eq:62} \ ,
\end{eqnarray}
where $\mathcal{B}=\sqrt{\beta^2 +4 \theta_E^2}$ \ .

Now we can calculate expressions for the Einstein angle by putting $\beta=0$ into Eqs (\ref{eq:60}), (\ref{eq:61}) and (\ref{eq:62}). We get the relations for the three images in the form 
\begin{eqnarray}
\theta_0^{(1)} &=& \theta_E-\frac{\theta_F}{2 \theta_E^4} \label{eq:63} \ , \\
\theta_0^{(2)} &=& -\theta_E-\frac{\theta_F}{2 \theta_E^4} \label{eq:64} \ , \\
\theta_0^{(3)} &=& \left(\frac{\theta_F}{\theta_E^2} \right)^{\frac{1}{3}} \label{eq:65} \ .
\end{eqnarray}
\begin{figure*}[t!]
\begin{center}
\includegraphics[width=0.3\linewidth]{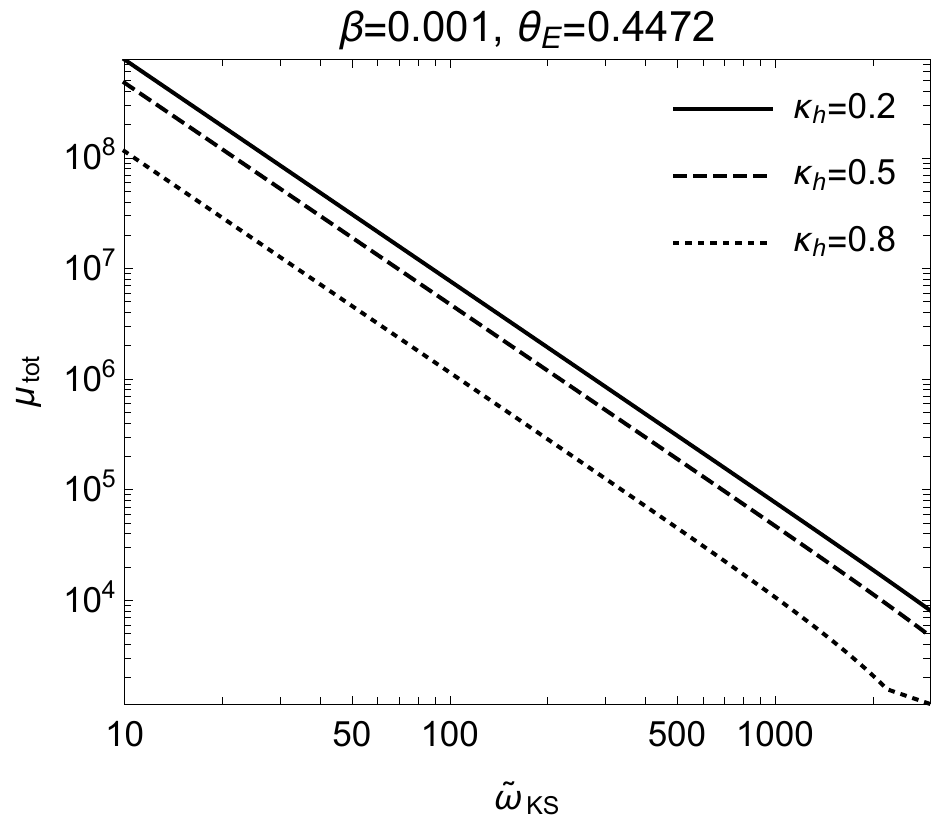} 
\includegraphics[width=0.3\linewidth]{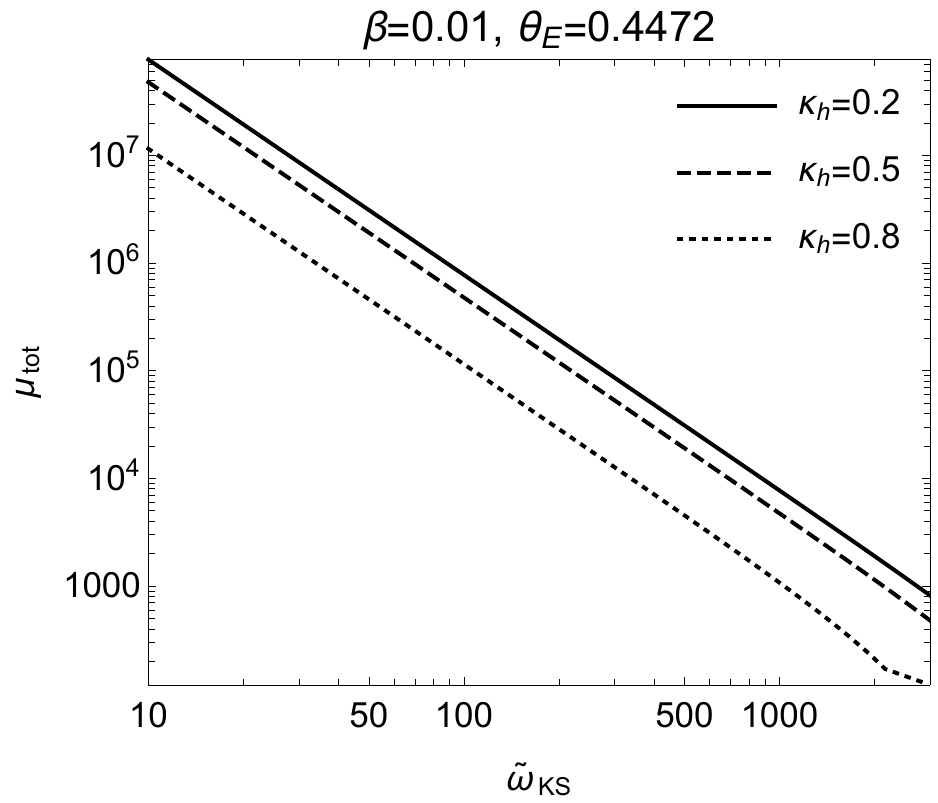}
\includegraphics[width=0.3\linewidth]{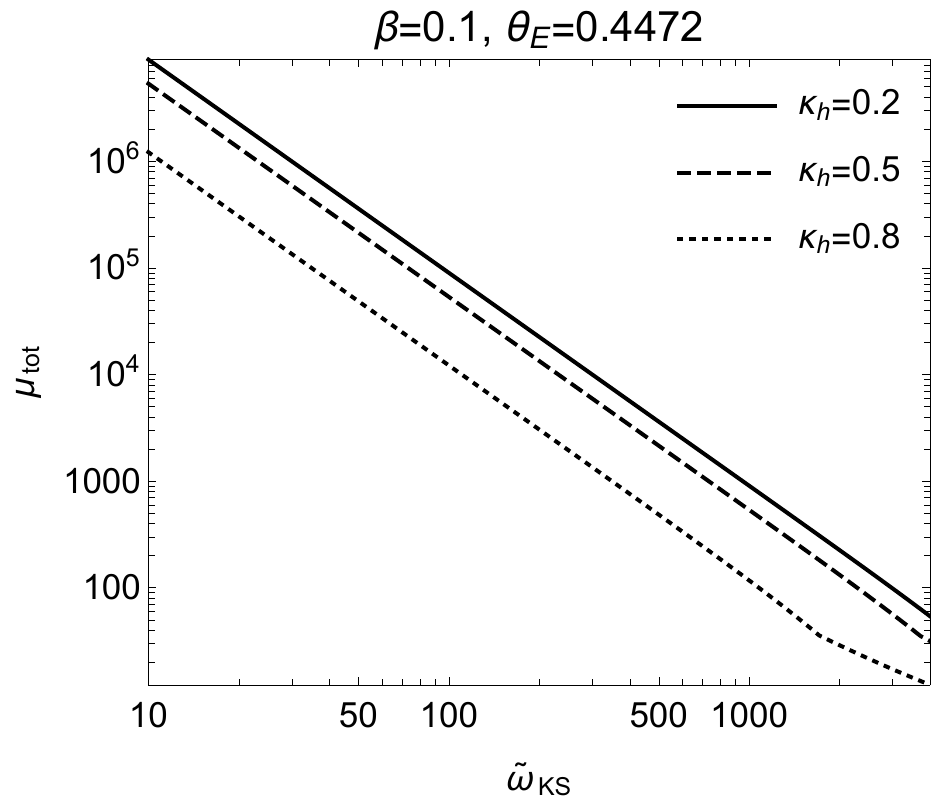}

\includegraphics[width=0.3\linewidth]{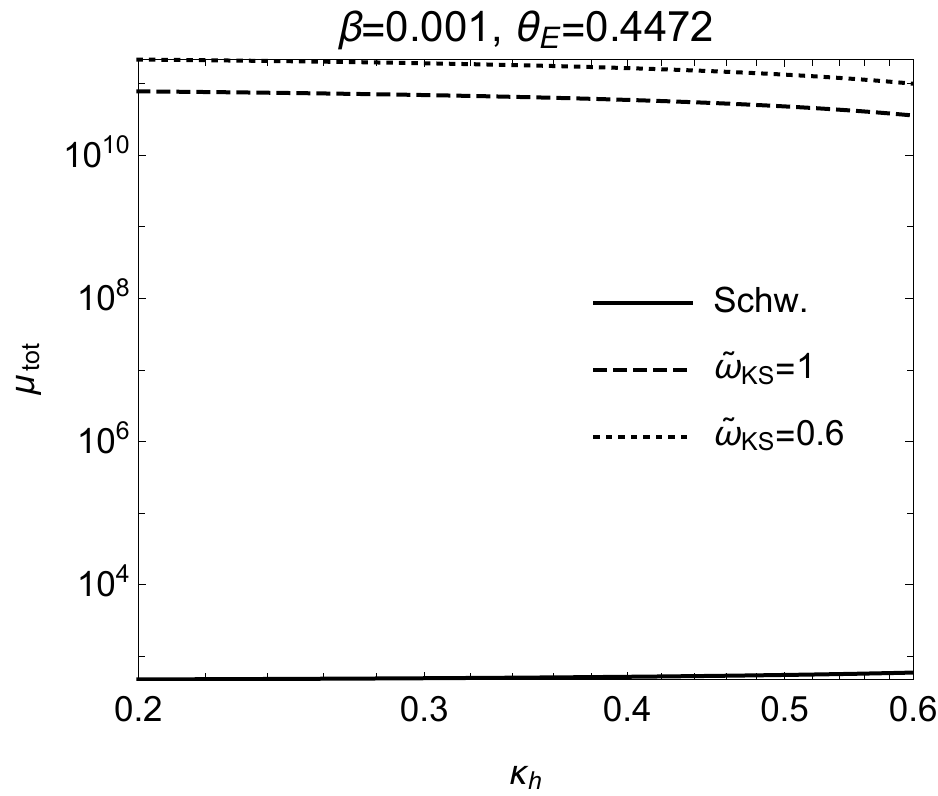} 
\includegraphics[width=0.3\linewidth]{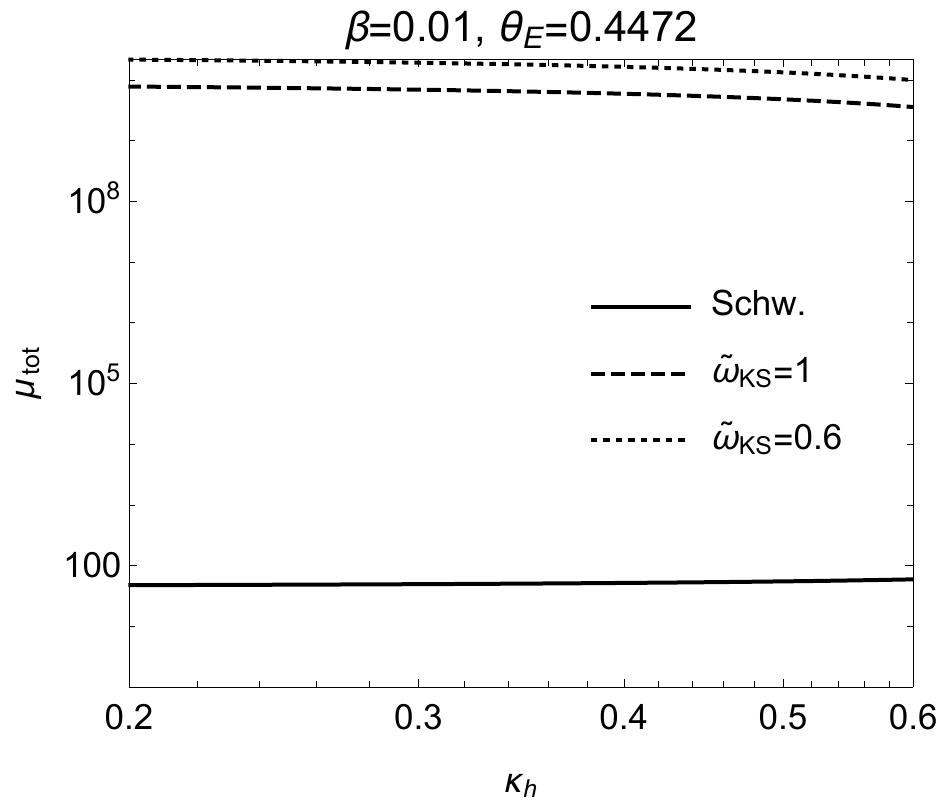}
\includegraphics[width=0.3\linewidth]{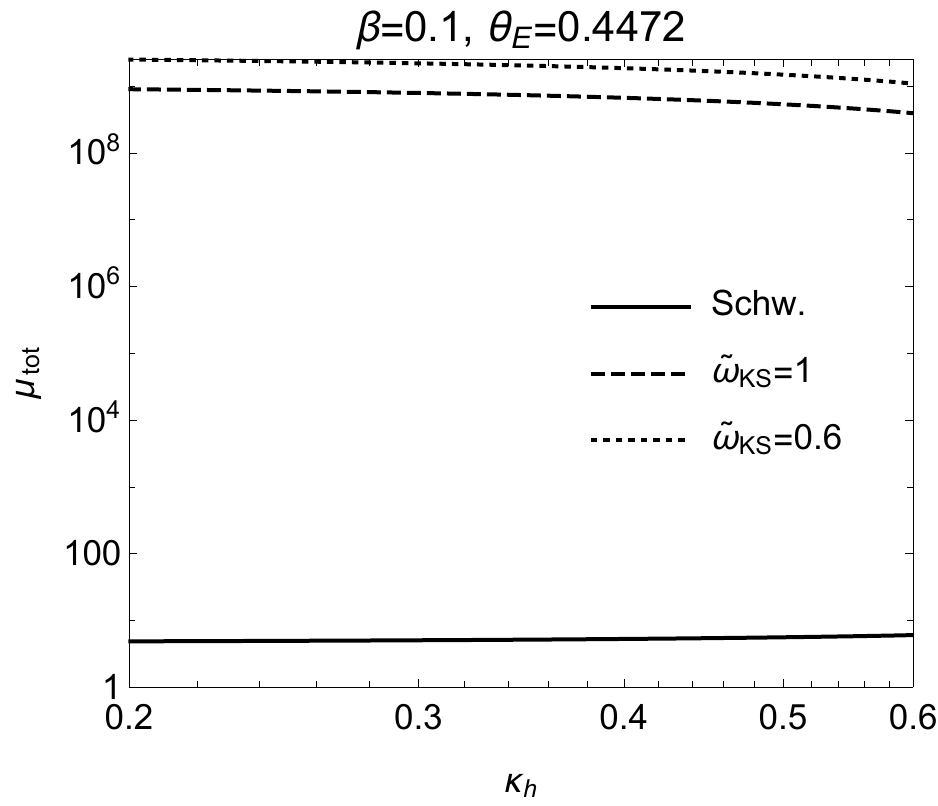} 
\end{center}
\caption{\textbf{Homogeneous case:} In the first row, the variation of total magnification~($\mu_{tot}$) for homogeneous distribution of plasma with dimensionless ``Ho\v rava'' parameter~($\tilde{\omega}_{_{KS}}$) for different combinations of angle of source from observer-lens axis~($\beta$) and plasma parameter keeping $\theta_E=0.4472$ constant.
In the second row, the variation of total magnification~($\mu_{tot}$) for homogeneous distribution of plasma with plasma parameter for different combinations of angle of source from observer-lens axis~($\beta$) and dimensionless ``Ho\v rava'' parameter~($\tilde{\omega}_{_{KS}}$) keeping $\theta_E=0.4472$ constant.   \label{uniformmag}}
\end{figure*}
The total magnification of the image brightness can be calculated by using Eq.~(\ref{eq:55}). In Fig~\ref{Einsangltotmag}, on the left panel, we demonstrate for the three Einstein angles their dependence on the dimensionless ``Ho\v rava'' parameter. At the Schwarzschild limit~(at high values of $\omega_{_{KS}}$), the upper two Einstein angles coincide to one, and lower one goes to zero. 
In Fig~\ref{Einsangltotmag}, on the right panel, we demonstrate variation of the total magnification of the image brightness with the dimensionless ``Hora\v va parameter", for various values of the inclination angle of source from the observer-lens axis ($\beta$). We see that the magnification is increasing with the inclination angle decreasing. 

%

\subsection{Image magnification in homogeneous plasma}
%
\begin{figure*}[t!]
\begin{center}
\includegraphics[width=0.3\linewidth]{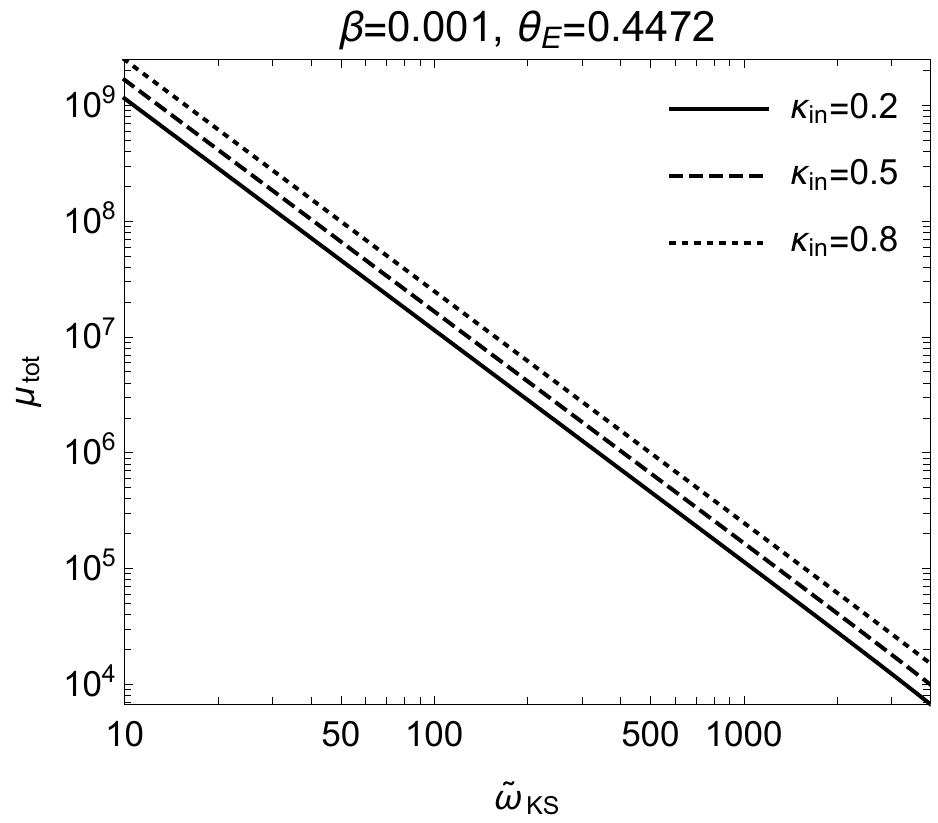} 
\includegraphics[width=0.3\linewidth]{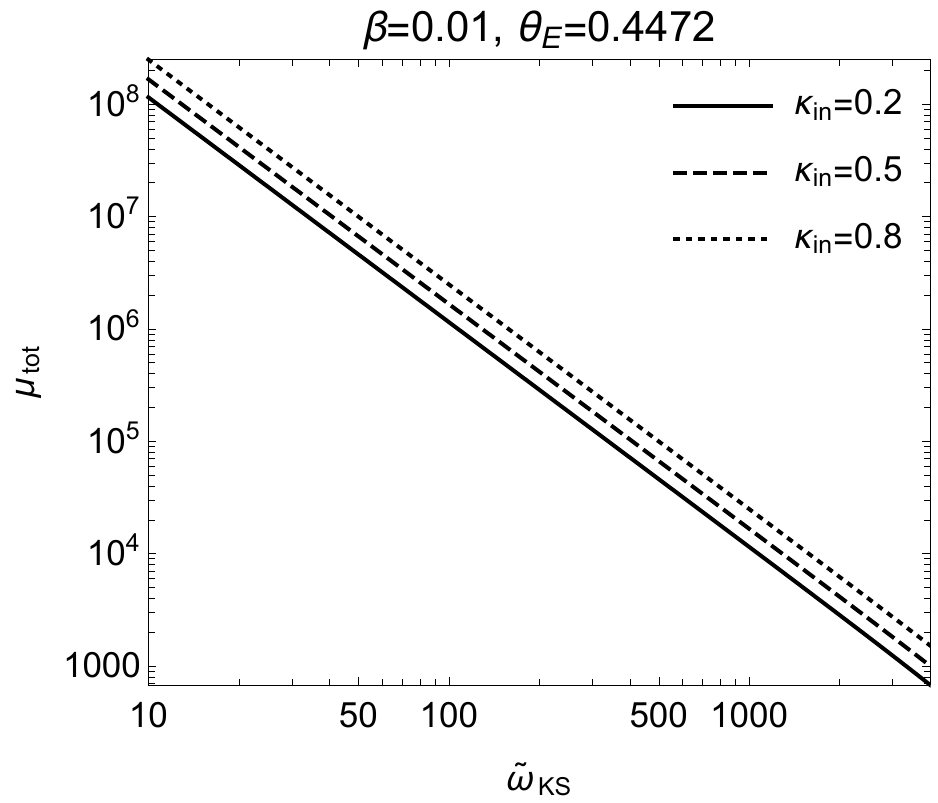}
\includegraphics[width=0.3\linewidth]{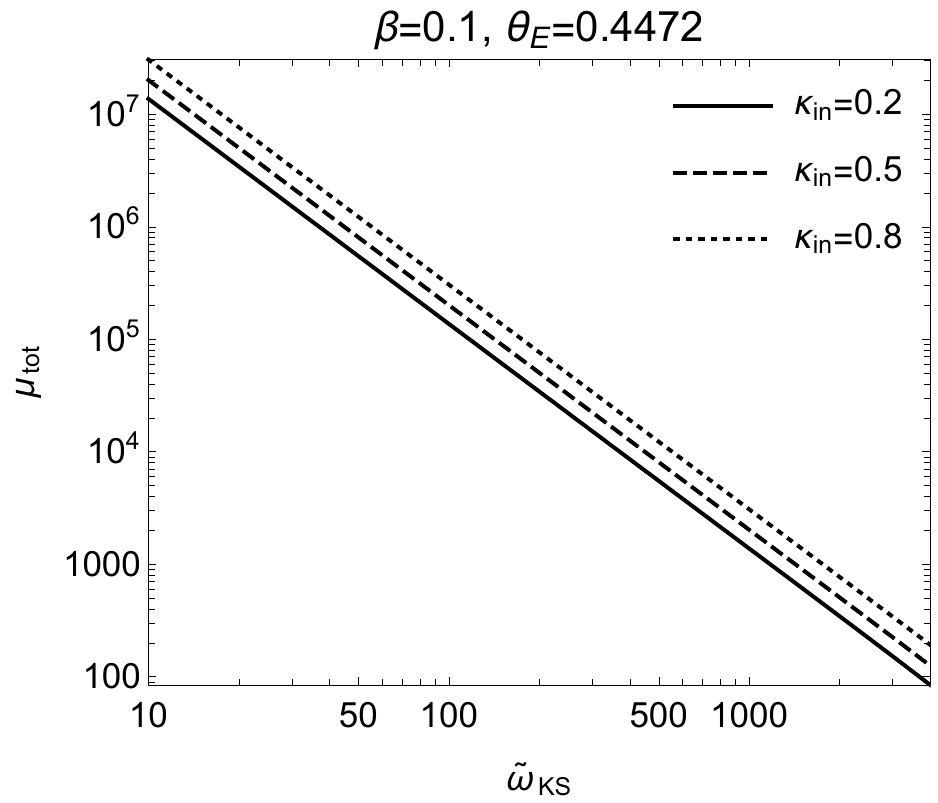}

\includegraphics[width=0.3\linewidth]{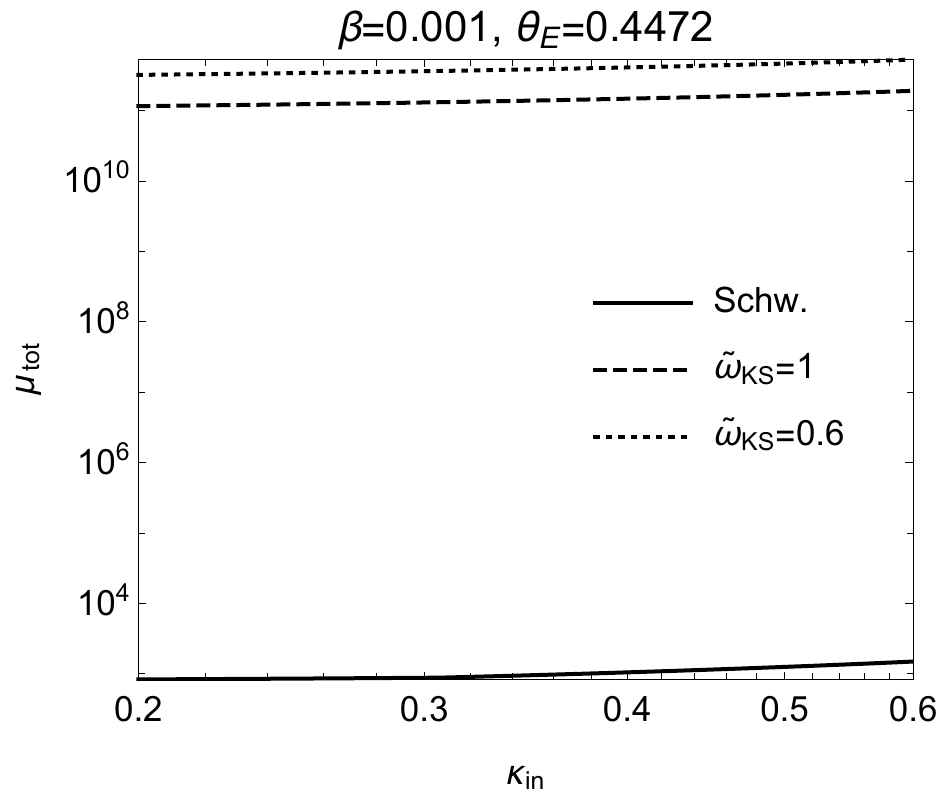} 
\includegraphics[width=0.3\linewidth]{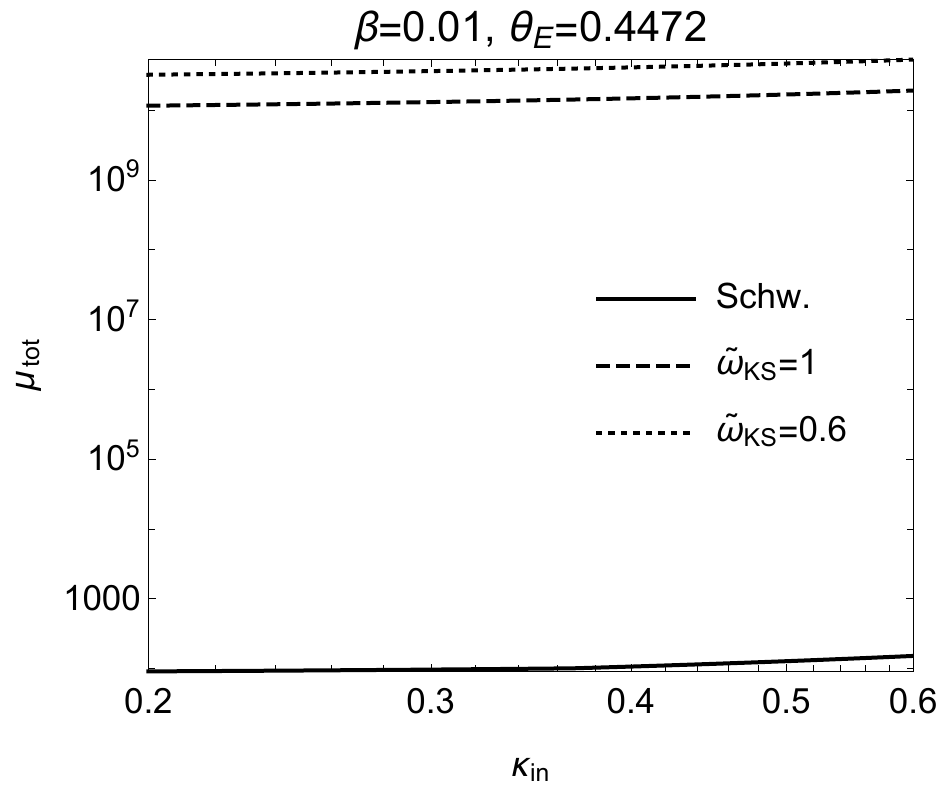}
\includegraphics[width=0.3\linewidth]{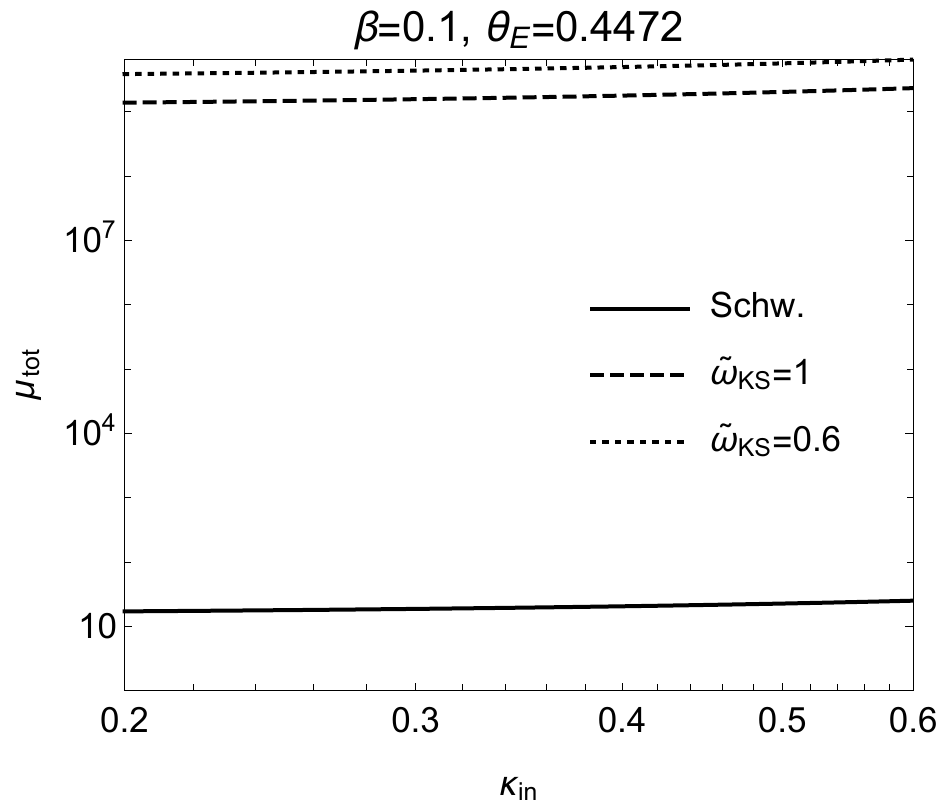} 
\end{center}
\caption{\textbf{Inhomogeneous:} In the first row, the variation of total magnification~($\mu_{tot}$) for inhomogeneous distribution~($1/r$) of plasma with dimensionless ``Ho\v rava'' parameter~($\tilde{\omega}_{_{KS}}$) for different combinations of angle of source from observer-lens axis~($\beta$) and plasma parameter keeping $\theta_E=0.4472$ constant.
In the second row, the variation of total magnification~($\mu_{tot}$) for inhomogeneous distribution~($1/r$) of plasma with plasma parameter for different combinations of angle of source from observer-lens axis~($\beta$) and dimensionless ``Ho\v rava'' parameter~($\tilde{\omega}_{_{KS}}$) keeping $\theta_E=0.4472$ constant. \label{powermag}}
\end{figure*}
By using Eqs~(\ref{eq:50})~and~(\ref{eq:54}), neglecting the terms involving~$\omega_{_{KS}}^2$, we obtain the lens equation in the case of homogeneous distribution of plasma in the form 
\begin{eqnarray}\label{eq:66}
 \theta^5-\beta \theta^4 - \frac{1}{2} \left(1+\frac{1}{1-\kappa_h}\right) \theta_E^2 \theta^3 \nonumber \\
 +\frac{4}{5} \left(\frac{1}{4}+\frac{1}{1-\kappa_h} \right) \theta_F=0 \ ,
\end{eqnarray}
where, $\theta_E$ and $\theta_F$ are given by Eqs~(\ref{eq:57}) and (\ref{eq:58}), respectively. We get three real roots corresponding to three different image positions by solving Eq.~(\ref{eq:66}). The solutions are given by 
\begin{eqnarray}
\theta_1 &=&\frac{1}{2} \left[\beta+\mathcal{C}+ \frac{16(-5 +\kappa_h)\theta_F}{5(1-\kappa_h) \mathcal{C}(\beta + \mathcal{C})^3} \right] \label{eq:67} \ , \\
\theta_2 &=&\frac{1}{2} \left[\beta-\mathcal{C}- \frac{16(-5 +\kappa_h)\theta_F}{5(1-\kappa_h) \mathcal{C}(\beta - \mathcal{C})^3} \right] \label{eq:68} \ , \\
\theta_3 &=& \left[\frac{8(\frac{1}{4}+\frac{1}{1-\kappa_h})\theta_F}{5 (1+\frac{1}{1-\kappa_h})\theta_E^2} \right] ^{\frac{1}{3}} \label{eq:69} \ ,
\end{eqnarray}
where, $\mathcal{C}=\sqrt{\beta^2 + 2 \theta_E^2(1+\frac{1}{1-\kappa_h})}$ \ .
We can obtain the expressions for the Einstein angle in similar way as in the vacuum case, by putting $\beta=0$ in the above expressions. 

In the left panel of Fig.~\ref{unipwr}, we present for homogeneous distribution of plasma the dependence of the Einstein angle($\theta_0$) on dimensionless ``Ho\v rava'' parameter for different values of $\theta_E$. We can see three Einstein rings for each value of $\theta_E$, when the dimensionless ``Ho\v rava'' parameter is low enough. As in the vacuum case, we obtain only one Einstein ring in the Schwarzschild limit( at large values of $\tilde{\omega}_{_{KS}}$). 
The total magnification of the image brightness can be calculated by using the Eq.~(\ref{eq:55}). In the first row of Fig~\ref{uniformmag}, we present the variation of the total magnification of the image brightness with the dimensionless ``Ho\v rava'' parameter, and with the plasma parameter, for various values of $\beta$. From these plots we observe that the total magnification is decreasing if we increase plasma parameter, and the total magnification is increasing if we decrease $\beta$. This means, the total magnification decreases if plasma concentration gets denser. From the second row of Fig~\ref{uniformmag}, we can conclude that the total magnification is increasing if the dimensionless ``Ho\v rava'' parameter decreases. So, the total magnification increases with strength of the KS spacetime. 
%
\subsection{Image magnification in inhomogeneous plasma}

%
We get the lens equation by using Eqs~(\ref{eq:54})(\ref{eq:52}) -- neglecting terms involving $\omega_{_{KS}}^2$) we arrive at
\begin{eqnarray} \label{eq:70}
\theta^5-\beta \theta^4- \left[\theta_E^2-\frac{ \theta_E^2 \kappa_{in}}{4M}\right] \theta^3 +\theta_F=0 \ ,
\end{eqnarray}
We get three images whose positions are given by 
\begin{eqnarray}
\theta_1 &=&\frac{1}{2} \left[\beta+\mathcal{H}-\frac{16 \theta_F}{ \mathcal{H}(\beta + \mathcal{H})^3} \right] \label{eq:71} \ , \\
\theta_2 &=&\frac{1}{2} \left[\beta-\mathcal{H}+ \frac{16 \theta_F}{ \mathcal{H}(\beta - \mathcal{H})^3} \right] \label{eq:72} \ , \\
\theta_3 &=& \left[\frac{4 \theta_F}{4 \theta_E^2-\frac{ \theta_E^2 \kappa_{in}}{M}} \right] ^{\frac{1}{3}} \label{eq:73} \ ,
\end{eqnarray}
where $\mathcal{H}=\sqrt{\beta^2 +4 \theta_E^2-\frac{\theta_{E}^2 \kappa_{in}}{M}}$ \ .

We can obtain the Einstein angles by setting $\beta=0$ in the above expressions. In the left panel of Fig.~\ref{unipwr}, we show for inhomogeneous distribution of plasma the dependence of the Einstein angle~($\theta_0$) on the dimensionless ``Ho\v rava'' parameter for different values of $\theta_E$. This case is similar to the case of homogeneous distribution of plasma. We can see three Einstein rings for each value of $\theta_E$, if the dimensionless ``Ho\v rava'' parameter is low enough, and we see only one Einstein ring at the Schwarzschild limit~(for large values of $\tilde{\omega}_{_{KS}}$). 

The total magnification of the image brightness can be calculated by using Eq.~(\ref{eq:55}). In the Fig.~\ref{powermag}, we demonstrate the variations of total magnification of the image brightness with the dimensionless ``Hora\v va parameter" and with the plasma parameter, for various values of $\beta$. In this case, the profile has similar character as in the homogeneous case, but the magnification increases significantly due to inhomogeneity in the plasma distribution.


\section{Conclusion\label{Summary}}

In this work we explore the optical properties of the KS black holes in the presence of plasma. We have studied the photon motion using the Hamilton-Jacobi equation, modified due to the presence of plasma. 

The radius of the shadow of the black hole is increasing when both the ``Ho\v rava'' parameter and the plasma parameter increase. It was shown that ``Ho\v rava'' parameter and the plasma parameter have significant influence on photons deflected near the compact object. In case of inhomogeneous distribution of plasma, the deflection angle is lower than in the homogeneous case, and it decreases monotonically with increase of plasma parameter. 

We also see that the image magnification is increasing when both the inlination angle of the source from the observer-lens axis and the dimensionless ``Ho\v rava'' parameter decrease. We can also conclude that if plasma concentration gets denser, the total magnification decreases. In case of injection of an inhomogeneity in the plasma distribution, the magnification increases.

Now we have the observational data of the Event Horizon Telescope~(EHT)~\cite{2019ApJ...875L...1E,2019ApJ...875L...2E,2019ApJ...875L...3E,2019ApJ...875L...4E,2019ApJ...875L...5E,2019ApJ...875L...6E} which we could use to verify results of our theoretical models with the observational data. If this or any future observation will match with our theoretical findings of deflection angle, magnification, number of image, radius of shadow then we can comment that the compact object is a Kehagias-Sfetsos black hole, and we could put limit on the ``Ho\v rava'' parameter. We could also be able to distinguish the effects of ``Ho\v rava'' parameter and plasma on the optical phenomena.
Next, we plan to study the effect of ``Ho\v rava'' parameter on the optical phenomena as gravitational lensing in strong gravitational field with anisotropic distribution of plasma. 
 
\section*{Acknowledgement}
S.H., J.S. and Z.S. would like to express their acknowledgements for the Institutional support of the Faculty of
Philosophy and Science of the Silesian University in
Opava, the internal student grant of the Silesian University SGS/12/2019 and the Albert Einstein Centre for
Gravitation and Astrophysics supported by the Czech
Science Foundation grant No. 14-37086G. A.A.
acknowledge the Faculty of Philosophy and Science,
Silesian University in Opava, Czech Republic and the
Goethe University, Frankfurt am Main, Germany for
their warm hospitality. This research is supported by by Grants No.~VA-FA-F-2-008 and No. YFA-Ftech-2018-8 of the Uzbekistan Ministry for Innovation Development, and by the
Abdus Salam International Centre for Theoretical
Physics through Grant No.~OEA-NT-01.
This research is partially supported by an
Erasmus+ exchange grant between SU and NUUz.
\bibliographystyle{apsrev4-1}  
\bibliography{gravreferences}

\end{document}